\documentclass[a4paper,12pt]{elsarticle}
\usepackage{slashed}
\usepackage{amssymb,amsmath,amsbsy}
\usepackage{mathrsfs}
\usepackage{dsfont}
\usepackage{bm}
\usepackage{epsfig}
\usepackage{hyperref} \usepackage{mdframed} \usepackage{adjustbox}
\usepackage{caption} \usepackage[makeroom]{cancel}
\usepackage[top=1.2in,bottom=1.2in,left=0.8in,right=0.8in,includefoot]{geometry}
\usepackage{xcolor} \allowdisplaybreaks \biboptions{sort&compress}
\usepackage[normalem]{ulem}
  \def\sfr{{\tt
    SmeftFR}~} \def\frules{{\tt FeynRules}~}
\def\webpage{\url{www.fuw.edu.pl/smeft}}
 \newcommand{\bea}{\begin{eqnarray}}
\newcommand{\eea}{\end{eqnarray}} \newcommand{\nn}{\nonumber\\}

\newcommand{\bce}{\begin{center}} \newcommand{\ece}{\end{center}}
\newcommand{\be}{\begin{equation}} \newcommand{\ee }{\end{equation}}
\newcommand{\btb}{\begin{tabular}} \newcommand{\etb}{\end{tabular}}

\def\lcal{{\cal L}}

\usepackage{color} 
\definecolor{orange}{rgb}{0.9,0.2,0}
\definecolor{brown}{rgb}{0.7,0.3,0.2} \definecolor{fuxia}{rgb}{1,0,1}
\definecolor{skyblue}{rgb}{0,0.1,0.9}
\definecolor{violetred}{rgb}{0.8,0.13,0.56}
\definecolor{deeppink}{rgb}{1.00,0.08,0.5}
\definecolor{pink}{rgb}{1.00,0.75,0.80}
\definecolor{orchid}{rgb}{0.85,0.44,0.84}
\definecolor{lightpink}{rgb}{1.00,0.71,0.76}
\definecolor{bluish}{rgb}{0,0.6,0.8}
\definecolor{lightgray}{rgb}{0.95,0.95,0.95}

\newcommand{\red}[1]{\color{red} #1 \color{black}}

\numberwithin{equation}{section}

\journal{Computer Physics Communications}

\begin{document}

\begin{frontmatter}

\title{\sfr v3 -- Feynman rules generator for the Standard Model
  Effective Field Theory}

\author[a]{A. Dedes} \author[b]{J. Rosiek\corref{author}}
\author[b]{M. Ryczkowski} \author[a]{K. Suxho}
\author[a]{L. Trifyllis}

\cortext[author]{Corresponding author, e-mail address
  \href{mailto:janusz.rosiek@fuw.edu.pl}{\tt
    janusz.rosiek@fuw.edu.pl}}

\address[a]{Department of Physics, University of Ioannina, GR 45110,
  Ioannina, Greece}

\address[b]{Faculty of Physics, University of Warsaw, Pasteura 5,
  02-093 Warsaw, Poland}

\begin{abstract}
  We present version 3 of {\tt SmeftFR}, a Mathematica package
  designed to generate the Feynman rules for the Standard Model
  Effective Field Theory (SMEFT) including the complete set of gauge
  invariant operators up to dimension-6 and the complete set of
  bosonic operators of dimension-8.  Feynman rules are generated with
  the use of {\tt FeynRules} package, directly in the physical (mass
  eigenstates) basis for all fields.  The complete set of interaction
  vertices can be derived, including all or any chosen subset of SMEFT
  operators.  As an option, the user can also choose preferred gauge
  fixing, generating Feynman rules in unitary or $R_\xi$-gauges.  The
  novel feature in version-3 of {\tt SmeftFR} is its ability to
  calculate SMEFT interactions consistently up to dimension-8 in EFT
  expansion (including quadratic dimension-6 terms) and express the
  vertices directly in terms of user-defined set of input-parameters.
  The derived Lagrangian in the mass basis can be exported in various
  formats supported by {\tt FeynRules}, such as {\tt UFO}, {\tt
    FeynArts}, \textit{etc}.  Initialisation of numerical values of
  Wilson coefficients of higher dimension operators is interfaced to
  WCxf format.  The package also includes a dedicated Latex generator
  allowing to print the result in clear human-readable form.  The {\tt
    SmeftFR} v3 is publicly available at \webpage.
\end{abstract}


\end{frontmatter}

\newpage

\noindent {\bf PROGRAM SUMMARY}

\begin{small}
\noindent
{\em Manuscript Title:} \\
\sfr v3 -- Feynman rules generator for the Standard Model Effective
Field Theory\\
{\em Authors:} A.  Dedes, J.  Rosiek, M.  Ryczkowski, K.  Suxho, L.
Trifyllis\\
{\em Program Title:} \sfr v3.0 \\
{\em Journal Reference:} \\
%Leave blank, supplied by Elsevier.
{\em Catalogue identifier:} \\
%Leave blank, supplied by Elsevier.
{\em Licensing provisions:} None \\
%enter "none" if CPC non-profit use license is sufficient.
{\em Programming language:} Mathematica 12.1 or later (earlier
versions were reported to have problems running this code) \\
{\em Computer:} any running Mathematica \\
{\em Operating system:} any running Mathematica \\
{\em RAM:} allocated dynamically by Mathematica, at least 4GB total
RAM suggested \\
{\em Number of processors used:} allocated dynamically by Mathematica
\\
{\em Supplementary material:} None 
{\em Classification:}\\ \begin{tabular}{ll}
11.1 & General, High Energy Physics and Computing,\\
4.4 & Feynman diagrams,\\
5 & Computer Algebra.  \\
\end{tabular}\\
%Classify using CPC Program Library Subject Index, see (
% http://cpc.cs.qub.ac.uk/subjectIndex/SUBJECT_index.html) e.g.   4.4
% Feynman diagrams, 5 Computer Algebra.
{\em External routines/libraries:} Wolfram Mathematica program \\
{\em Subprograms used:} {\tt FeynRules v2.3.49} or later package \\
{\em Nature of problem:}\\
Automatised generation of Feynman rules in physical (mass) basis for
the Standard Model Effective Field Theory with user defined operator
subset, gauge fixing and input-parameters scheme selection.  \\
{\em Solution method:}\\
Expansion of \sfr v2 Mathematica package~[1]: implementation of the
results of Ref.~[2] in the \frules package~[3], including dynamic
``model files'' generation.  \\
{\em Restrictions:} None\\
%Describe any restrictions on the complexity of the problem here.
{\em Unusual features:} None \\
%Describe any unusual features of the program/problem here.
{\em Additional comments:} None \\
{\em Running time:} depending on control variable settings, from few
minutes for a selected subset of few SMEFT operators and Feynman rules
generation up to several hours for generating UFO output for large
operator sets (using Mathematica 13.2 running on a personal computer)
\\

\end{small}

\newpage

%%%%%%%%%%%%%%%%%%%%%%%%%%%%%%%%%
\section{Introduction}
\label{sec:intro}

The Standard Model Effective Field Theory
(SMEFT)~\cite{Weinberg:1980wa,Coleman:1969sm,Callan:1969sn} is a
useful tool in parameterizing phenomena beyond the, successful so far,
Standard Model (SM)~\cite{Weinberg:1967tq, Glashow, Salam} predictions
that may appear in current and/or future particle experiments.  The
SMEFT Lagrangian is given by
%%%%%%%%%%%%%%%
\begin{equation}
\mathcal{L}_{\mathrm{SMEFT}} \ = \ \mathcal{L}_{\mathrm{SM}} \ +
\ \sum_i \frac{C_i \, \mathcal{O}_i}{\Lambda^{d_i-4}} \;,
\label{eq:1}
\end{equation}
 %%%%%%%%%%%%%%
where scale $\Lambda$ is the cut-off scale of the EFT (i.e., the mass
of the lightest heavy particle decoupled from the underlying theory),
$\mathcal{O}_i$ is a set of $d_i$-dimensional, SM gauge group
invariant, operators, and $C_i$ are the associated Wilson coefficients
(WCs).  For one fermion generation including Hermitian conjugation, we
have for $d=5$ two independent operators e.g.  $i=2$, for $d=6$ we
have $i=84$, for $d=7$ we have $i=30$, for $d=8$ we have $i=993$,
%for $d=9$ we have $i=560$, for $d=10$ we have $i=15456$,
and so on and so forth~\cite{Henning:2015alf}.  When expanding in
flavour, the actual number of operators explodes from few to several
thousand of operators and interaction vertices.  This proliferation of
vertices must be included in matrix element calculators when mapping
the WCs to experimental data.  This is the scope of this article: to
describe the code \sfr v3.0 which consistently provides the Feynman
Rules for dimension-6 and the bosonic part of dimension-8 operators
for further symbolic or numerical manipulations.

Admittedly, SMEFT is a hugely complicated model.  Including all
possible CP-, flavour-, baryon-, and lepton-number violating
interactions at dimension-6 level, it already contains 2499 free
parameters in a non-redundant basis, such as the Warsaw
basis~\cite{Grzadkowski:2010es}.  In addition, recent experimental and
theoretical progress of high energy processes at LHC involving vector
boson scattering requires subsets of dimension-8
operators~\cite{Murphy:2020rsh, Li:2020gnx}, in particular the bosonic
ones, making the structure of possible interactions even more
involved.  Due to large number and complicated structure of new terms
in the Lagrangian, theoretical calculations of physical processes
within the SMEFT can be very challenging --- it is enough to notice
that the number of primary vertices when SMEFT is quantized in
$R_\xi$-gauges and in ``Warsaw mass'' basis, printed for the first
time in ref.~\cite{Dedes:2017zog}, is almost 400 without counting the
hermitian conjugates.

As a result, it is important to develop technical methods and tools
facilitating such calculations, starting from developing the universal
set of the Feynman rules for propagators and vertices for physical
fields, after Spontaneous Symmetry Breaking (SSB) of the full
effective theory in the most commonly studied, Warsaw basis of
operators~\cite{Grzadkowski:2010es}. The initial version of the
relevant package, \sfr v1.0, was announced and briefly described for
the first time in Appendix B of ref.~\cite{Dedes:2017zog}.  The \sfr
code was further developed and supplied with many new options
capabilities and published as \sfr v2.0 in~\cite{Dedes:2019uzs}.  In
this paper we present \sfr v3.0, a {\em Mathematica} symbolic language
package generating Feynman rules in several formats, based on the
formulae developed in ref.~\cite{Dedes:2017zog}.  The most important
new capability implemented in the code, comparing to version 2, allows
for performing consistent calculations up to dimension-8 operators in
EFT expansion, including also expressing the Feynman rules directly in
terms of any user-defined set of input parameters.  We summarise here
the main features of \sfr code, noting in particular advances
introduced in its 3rd version (v3):

\begin{itemize}
\item \sfr is written as an overlay to \frules
  package~\cite{Christensen:2008py, Alloul:2013bka}, used as the
  engine to generate Feynman rules.
\item \sfr v3 is able to generate interactions in the most general
  form of the SMEFT Lagrangian up to dimension-6 order in Warsaw
  basis~\cite{Grzadkowski:2010es}, without any restrictions on the
  structure of flavour violating terms and on CP-, lepton- or
  baryon-number conservation.  In addition, it also contains all {\em
    bosonic} operators of dimension-8 order, in the basis defined in
  ref.~\cite{Murphy:2020rsh}.
\item Feynman rules are expressed in terms of physical SM fields and
  canonically normalised Goldstone and ghost fields.  Expressions for
  interaction vertices are analytically expanded in powers of inverse
  New Physics scale $1/\Lambda$.  The novel feature implemented in
  \sfr v3 is the consistent inclusion of all terms up to maximal
  dimension-8, including both terms quadratic in Wilson coefficients
  of dimension-6 and linear contributions from Wilson coefficients of
  dimension-8.  Terms of order higher than $d=8$ are consistently
  truncated.
\item Another important novel feature of \sfr v3 is the possibility of
  expressing Feynman rules directly in terms of a predefined set of
  input parameters (usually chosen to be observables directly
  measurable in experiments).  This allows for consistent calculation
  of processes in SMEFT without the complicated and error-prone
  procedure of using ``intermediate'' set of Lagrangian parameters and
  later re-expressing them in terms of preferred input quantities.
\item {\sfr} v3 allows for choosing {\em any} set of input parameters,
  assuming that the user provides appropriate routines relating them
  to ``standard'' SM Lagrangian parameters (defined later in
  Sec.~\ref{sec:pars}) to a required (maximum 8th) order of SMEFT
  expansion.  Two most frequently used input schemes in the
  electroweak sector, $(G_F, M_Z, M_W, M_H)$ and $(\alpha_{em}, M_Z,
  M_W, M_H)$ are predefined in current version, including all terms up
  to dimension-8.  In both cases, the strong coupling constant and all
  quark and lepton masses are also inputs.  In addition, \sfr v3 also
  includes a predefined input scheme for the CKM matrix adopted from
  ref.~\cite{Descotes-Genon:2018foz}.  For the neutrino mixing matrix
  we use as input the standard PMNS matrix not (as yet) corrected by
  SMEFT.
\item Including the full set of SMEFT parameters in model files for
  \frules may lead to very slow computations.  \sfr can generate
  \frules model files dynamically, including only the user defined
  subset of higher dimension operators.  It significantly speeds up
  the calculations and produces a simpler final result, containing
  only the Wilson coefficients relevant to the process that she/he has
  chosen to analyse.  It is worth noting that optimisations included
  in \sfr v3 sped it up comparing to \sfr v2 by approximately an order
  of magnitude for a comparable subset of chosen operators of
  dimension-6 and calculations done up to $1/\Lambda^2$ accuracy
  (maximally achievable in \sfr v2).
\item Feynman rules can be generated in the unitary or in linear
  $R_\xi$-gauges by exploiting four different gauge-fixing parameters
  $\xi_\gamma, \xi_Z, \xi_W, \xi_G$ for thorough amplitude checks.  In
  the latter case also all relevant ghost and Goldstone vertices are
  obtained.  This procedure is described in detail in
  ref.~\cite{Dedes:2017zog} and implemented already in \sfr
  v2~\cite{Dedes:2019uzs}.
\item Feynman rules are calculated first in {\em Mathematica}/\frules
  format.  They can be further exported in other formats: {\tt
    UFO}~\cite{Degrande:2011ua} (importable to Monte-Carlo generators
  like {\tt MadGraph5\_aMC@NLO 5} ~\cite{Alwall:2014hca}, {\tt
    Sherpa}~\cite{Gleisberg:2008ta}, {\tt
    CalcHEP}~\cite{Belyaev:2012qa}, {\tt Whizard}~\cite{Kilian:2007gr,
    Christensen:2010wz}), {\tt FeynArts}~\cite{Hahn:2000kx} which
  generates inputs for loop amplitude calculators like {\tt
    FeynCalc}~\cite{Shtabovenko:2016sxi}, or {\tt
    FormCalc}~\cite{Hahn:2010zi}, and other output types supported by
  \frules.
\item \sfr provides a dedicated Latex generator, allowing to display
  vertices and analytical expressions for Feynman rules in clear human
  readable form, best suited for hand-made calculations.
\item \sfr is interfaced to the WCxf format~\cite{Aebischer:2017ugx}
  of Wilson coefficients.  Numerical values of SMEFT parameters in
  model files can be read from WCxf JSON-type input produced by other
  computer codes written for SMEFT.  Alternatively, \sfr can translate
  \frules model files to the WCxf format.
\item Further package options allow to treat neutrino fields as
  massless Weyl or (in the case of non-vanishing dimension-5 operator)
  massive Majorana fermions, to correct signs in 4-fermion
  interactions not yet fully supported by \frules and to perform some
  additional operations as described later in this manual.
\end{itemize}

It has also been made and tested to be compatible with many other
publicly available high-energy physics related computer codes
accepting standardised input and output data formats.

%%%%%%%%%%%%%%%%%5
Feynman rules derived in ref.~\cite{Dedes:2017zog} using the {\tt
  SmeftFR} package have been used successfully in many articles,
including refs.~\cite{Dedes:2018seb, Dedes:2019bew, Dawson:2018pyl,
  Vryonidou:2018eyv, Hesari:2018ssq, Dawson:2018liq, Dawson:2018jlg,
  Baglio:2018bkm, Dawson:2018dxp, Silvestrini:2018dos,
  Neumann:2019kvk, Arnan:2019uhr, Cullen:2019nnr, Aoude:2019tzn,
  Liu:2019wmi, Boughezal:2019xpp, Dawson:2019clf, Durieux:2019eor,
  Baglio:2020oqu, Dedes:2020xmo, Endo:2020kie, Muller:2021jqs,
  Anisha:2021fzf, Camargo-Molina:2021zgz, Cao:2021wcc, Dawson:2021ofa,
  Chen:2021rid, Dedes:2021abc, Atkinson:2021jnj, Alasfar:2022zyr,
  DiLuzio:2022xns, Bhardwaj:2022qtk, Asteriadis:2022ras,
  Boughezal:2023ooo, CMS:2022ztx}, and have passed certain non-trivial
tests, such as gauge-fixing parameter independence of the $S$-matrix
elements, validity of Ward identities, cancellation of infinities in
loop calculations, \textit{etc}.

We note, here, that there is a growing number of publicly available
codes performing computations related to
SMEFT~\cite{Proceedings:2019rnh}.  These include, {\tt
  Wilson}~\cite{Aebischer:2018bkb}, {\tt
  Flavio}~\cite{Straub:2018kue}, {\tt
  DSixTools}~\cite{Celis:2017hod,Fuentes-Martin:2020zaz}, {\tt
  RGESolver}~\cite{DiNoi:2022ejg}, {\tt
  MatchingTools}~\cite{Criado:2017khh}, {\tt
  CoDEx}~\cite{Bakshi:2018ics}, {\tt HighPT}~\cite{Allwicher:2022mcg},
{\tt STream}~\cite{Cohen:2020qvb}, {\tt
  SuperTracer}~\cite{Fuentes-Martin:2020udw}, {\tt
  Matchmakereft}~\cite{Carmona:2021xtq}, {\tt
  Matchete}~\cite{Fuentes-Martin:2022jrf}, which are codes for running
and matching Wilson coefficients and {\tt
  FeynOnium}~\cite{Brambilla:2020fla} for automatic calculations in
non-relativistic EFTs.  Packages mostly relevant to the purposes of
\sfr are {\tt SMEFTsim}~\cite{Brivio:2017btx,Brivio:2020onw}, {\tt
  Dim6Top}~\cite{Aguilar-Saavedra:2018ksv} and {\tt
  SMEFT@NLO}~\cite{Degrande_2021} which are all codes for calculating
physical observables in SMEFT.  To a degree, these codes (especially
the ones supporting WCxf format) can be used in conjunction with {\tt
  SmeftFR}.  For example, some of them can provide the numerical input
for Wilson coefficients of higher dimensional operators at scale
$\Lambda$, while others, the running of these coefficients from that
scale down to the EW one.  Alternatively, Feynman rules evaluated by
\sfr can be used with Monte-Carlo event generators to test the
predictions of other codes.

The rest of the paper is organised as follows.  In
Sec.~\ref{sec:basis}, we define the notation and conventions of the
SMEFT Lagrangian and the field normalisations used in transition to
mass basis.  In Sec.~\ref{sec:pars} and \ref{app:inp}, we describe the
input schemes, i.e. the user-defined choices of observables which can
be used to parametrize SMEFT interactions and give examples of the
corresponding output of the code.  In Sec.~\ref{sec:install}, we
present the code's algorithmic structure and installation procedure.
Sec.~\ref{sec:sfr} is the main part of the paper, illustrating in
detail how to derive the set of SMEFT vertices in mass basis starting
from $d=6$ operators in Warsaw basis~\cite{Grzadkowski:2010es} and
$d=8$ bosonic operators in basis of ref.~\cite{Murphy:2020rsh} (all
operators used by \sfr v3 are collected for completeness in
\ref{app:ops}).  A sample program with \sfr v3 commands, generating
Feynman rules in various formats, is given in Sec.~\ref{sec:sample}.
We conclude in Sec.~\ref{sec:summary}.
 
%%%%%%%%%%%%%%%%%%%%%%%%%%%%%%%
\section{SMEFT Lagrangian in gauge and mass basis}
\label{sec:basis}
%%%%%%%%%%%%%%%%%%%%%%%%%%%%%%%%%%%%

The classification of higher order operators in SMEFT is done in terms
of fields in electroweak basis, before Spontaneous Symmetry Breaking
(SSB).  For the dimension-5 and -6 operators, \sfr uses the so-called
``Warsaw basis''~\cite{Grzadkowski:2010es} as a starting point to
calculate physical states in SMEFT and their interactions (for the
specification of Warsaw basis, see ref.~\cite{Grzadkowski:2010es}, in
particular eq.~(3.1) defining the $d=5$ Weinberg operator
$Q_{\nu\nu}^{(5)}$ and Tables 2 and 3 containing the full list of
$d=6$ operators).  For the dimension-8 operators, we include all
operators containing bosonic fields only, as listed in Tables 2 and 3
of ref.~\cite{Murphy:2020rsh} with an exception of two operators.  The
definitions and the list of all operators used by \sfr v3 is described
in \ref{app:ops}, and
Tables~\ref{tab:sfr3-dim6},~\ref{tab:dim8-operators-higgs},
~\ref{tab:dim8-operators-gauge},
and~\ref{tab:dim8-operators-gauge-higgs}.

We decided to neglect $d=7$ (which always contain fermionic fields)
and fermionic $d=8$ operators, both for theoretical and practical
purposes.  Dimension-7 operators are all lepton or baryon number
violating and strongly constrained by many, related, experiments.  In
most BSM models, dimension-8 operators are also strongly suppressed
and can lead to substantial measurable effects only when their
contributions are enhanced, which typically (as can be justified on
dimensional ground) happens at high energies.  Such effects could be
in particular investigated in experimental searches that involve
Vector Boson Scattering at the LHC (see e.g.~\cite{Kalinowski:2018oxd,
  Doroba:2012pd, Dedes:2020xmo, BuarqueFranzosi:2021wrv,
  Covarelli:2021gyz}), and therefore, including bosonic operators is
particularly important for such contemporary analyses.  Furthermore,
fermionic $d=8$ operators, either pure or mixed with other fields, may
be equally important for collider studies. Chosen higher order
fermionic operators can also be loaded in \sfr, however, as we will
discuss in Section~\ref{sec:future}, at present it requires
introducing certain modifications and thus some expertise in the code
structure.

The SMEFT Lagrangian which we use is the sum of the dimension-4 terms
and operators of order up to dimension-8 (the latter only in the
bosonic sector):
\be
\lcal = \lcal_{\mathrm SM}^{(4)} + \frac{1}{\Lambda} C^{\nu\nu}
Q_{\nu\nu}^{(5)} + \frac{1}{\Lambda^2} \sum_{boson,fermion}
C_{(b,f)}^{(6)} Q_{(b,f)}^{(6)} + \frac{1}{\Lambda^4} \sum_{boson}
C^{(8)}_{b} Q_b^{(8)} \;.
\label{Leff}
\ee
Physical fields in SMEFT are obtained after SSB.  In the gauge and
Higgs sectors, physical and Goldstone fields ($h, G^0,G^\pm,
W^\pm_\mu, Z^0_\mu, A_\mu$) are related to initial (Warsaw basis)
fields ($\varphi, W_\mu^i, B_\mu, G_\mu^A$) by field normalisation
constants:\footnote{Note the notation difference with
ref.~\cite{Dedes:2017zog}: Quantities $Z_W$ and $Z_G$ defined in
eq.~\ref{eq:rot} are denoted as their inverses, $Z_W^{-1}$ and
$Z_G^{-1}$, in ref.~\cite{Dedes:2017zog}.}
%%%%%%%%%%%%%
\bea
\left( \begin{array}{c} \varphi^+ \\ \varphi^0 \end{array} \right) &=&
\left ( \begin{array}{c} Z_{G^+}^{-1} G^+ \\ \frac{1}{\sqrt{2}} (v +
  Z_h^{-1} h + i Z_{G^0}^{-1} G^0) \end{array} \right ) \;,\nn
\left( \begin{array}{c} {W}^3_\mu \\ {B}_\mu \end{array} \right) & =&
 Z_{\gamma Z} \left (\begin{array}{c} {Z}_\mu \\ {A}_\mu
  \end{array} \right ) \;,
\label{eq:rot}\nn
W^1_{\mu} &=& {Z_W\over \sqrt{2}}\,( W_\mu^+ + W_\mu^-)
\;, \label{Wpm-rot}\nn
W^2_{\mu} &=& {i Z_W\over \sqrt{2}}\,( W_\mu^+ - W_\mu^-)
\;, \label{Wpm-rot1}\nn
G_\mu^A &=& Z_{G} \, g_\mu^A \;.
\label{eq:hgnorm}
\eea
In addition, we define the effective gauge couplings, chosen to
preserve the natural form of covariant derivative:
\bea
\begin{array}{l}
g\, = Z_{g} \bar g \qquad
g' = Z_{g'} \bar g' \qquad
g_s = Z_{g_s} \bar g_s\,.
\end{array}
\label{Zg-norm} 
\eea
Up to $d=8$, the normalisation constants multiplying the gauge
couplings read as:
\bea
Z_{g} &=& \left( 1 - \frac{2v^2}{\Lambda^2} C_{\varphi W}
- \frac{v^4}{\Lambda^4} C_{W2\varphi 4n1} \right)^{1/2}\;, \\
Z_{g'} &=& \left( 1 - \frac{2v^2}{\Lambda^2} C_{\varphi B}
- \frac{v^4}{\Lambda^4} C_{B2\varphi 4n1} \right)^{1/2}\;, \\
 Z_{g_s} &=& \left( 1 - \frac{2v^2}{\Lambda^2} C_{\varphi G}
- \frac{v^4}{\Lambda^4} C_{G2\varphi 4n1} \right)^{1/2}\;,
\label{Zg-normdef} 
\eea
where relevant operators are defined in~\cite{Grzadkowski:2010es,
  Murphy:2020rsh} and formally all expressions have to be expanded to
the order $\frac{v^4}{\Lambda^4}$.

The above field normalisation constants $Z_X$, the corrected Higgs
field vev, $v$, and the gauge and Higgs boson masses, $M_Z$, $M_W$ and
$M_h$, are not encoded as fixed analytical expressions but calculated
by \sfr using the condition that bilinear part of the Lagrangian must
have canonical form in the mass eigenstates basis.  In this way, all
relations automatically contain only the subset of non-vanishing SMEFT
Wilson coefficients chosen by the user, as described in
Sec.~\ref{sec:sfr}.  The analytical expressions for the normalisation
constants for a chosen set of higher dimension operators after running
\sfr initialisation procedure are stored in variables listed in
Table~\ref{tab:zx} (as discussed later in Sec.~\ref{sec:pars},
expressions for the SM parameters in terms of user-defined input
quantities are also available, see Table~\ref{tab:userinput}).  One
should note that {\em at any order} in SMEFT, $SU(2)$ and $SU(3)$
gauge field and gauge normalisation constants are related,
$Z_W=Z_g^{-1}$, $Z_G=Z_{g_s}^{-1}$.

\begin{table}[t]
\begin{center}
\begin{tabular}{cp{2mm}cp{2mm}cp{2mm}c}
\hline  
Constant && Variable && Constant && Variable \\
\hline
$Z_{g_s}$ && {\tt gsnorm} && $Z_G$ && {\tt Gnorm} \\
$Z_{g}$ && {\tt gwnorm} && $Z_W$ && {\tt Wnorm} \\
$Z_{g'}$ && {\tt g1norm} && $Z_{\gamma Z}^{ij}$ && {\tt AZnorm[i,j]}
\\
$Z_{h}$ && {\tt Hnorm} && $Z_{G^0}$ && {\tt G0norm} \\
$Z_{G^+}$ && {\tt GPnorm} && &&  \\
\hline
\end{tabular}
\end{center}
\caption{Names of normalisation constants and corresponding internal
  \sfr variables.  \label{tab:zx}}
\end{table}

It is also easy to eventually further expand the program in future by
adding even higher than dimension-8 operators, as the routine
diagonalizing the field bilinears does not depend on their particular
dependence on Wilson coefficients of higher dimension operators until
the very final stage where such dependence is substituted and further
expanded in $1/\Lambda$ powers.

The basis in the fermion sector is not fixed by the structure of gauge
interactions and allows for unitary rotation freedom in the flavour
space:
\begin{equation}
\psi_X^\prime = U_{\psi_X} \: \psi_X\;,
\end{equation}
with $\psi=\nu,e,u,d$ and $X=L,R$.  We choose the rotations such that
$\psi_X$ eigenstates correspond to real and non-negative eigenvalues
of $3\times 3$ fermion mass matrices:
%%%%%%%%%%%%%%%
\begin{equation}
\begin{array}{cc}
M^\prime_\nu = - v^2 C^{\prime \nu\nu}\;, &
M^\prime_e = \frac{v}{\sqrt{2}}\, \left (\Gamma_e - \frac{v^2}{2}
C^{\prime e\varphi} \right ),\; \\[2mm]
M^\prime_u = \frac{v}{\sqrt{2}}\, \left (\Gamma_u - \frac{v^2}{2}
C^{\prime u\varphi} \right ),\; &
M^\prime_d = \frac{v}{\sqrt{2}}\, \left (\Gamma_d - \frac{v^2}{2}
C^{\prime d\varphi} \right ).
\end{array}
\label{eq:modyuk}
\end{equation}
The fermion flavour rotations can be adsorbed in redefinitions of
Wilson coefficients, as a result leaving CKM and PMNS matrices
(denoted in \sfr as $K$ and $U$ respectively) multiplying them.  The
complete list of redefinitions of flavour-dependent Wilson
coefficients is given in Table~4 of ref.~\cite{Dedes:2017zog}.  After
rotations, they are defined in so called ``Warsaw mass'' basis (as
also described in WCxf standard~\cite{Aebischer:2017ugx}).  \sfr
assumes that the numerical values of Wilson coefficients of $d=6$
fermionic operators (see Table~\ref{tab:sfr3-dim6}) are given in this
particular basis.

In summary, Feynman rules generated by the \sfr code describe
interactions of SMEFT physical (mass eigenstates) fields, with
numerical values of Wilson coefficients defined in the ``Warsaw mass''
basis of ref.~\cite{Dedes:2017zog} extended with bosonic subset of
dimension-8 operators in the basis defined in
ref.~\cite{Murphy:2020rsh}.

It is also important to stress that in the general case of lepton
number flavour violation, with the non-vanishing dimension-5 Weinberg
operator $Q^{(5)}_{\nu\nu}$, neutrinos are massive Majorana spinors,
whereas under the assumption of $L$-conservation they can be regarded
as massless Weyl spinors.  As described in the Sec.~\ref{sec:init},
\sfr is capable to generate Feynman rules for neutrino interactions in
both cases, depending on the choice of initial options\footnote{One
should remember that treating neutrinos as Majorana particles requires
special set of rules for propagators, vertices, and diagram
combinatorics.  We follow here the treatment described in
refs.~\cite{Denner:1992vza, Denner:1992me, Dedes:2017zog,
  Paraskevas:2018mks}.}. One should note that although for pure $V-A$
neutrino-gauge boson interactions in the SM the predictions for
physical observables almost never depend on the character of neutrino
fields (Dirac or Majorana), this is no longer true in case of
non-standard neutrino couplings generated by higher dimension
operators.  Detailed discussion of such issues, with relevant examples
of different predictions for 2- and 3-body decays involving pair of
Dirac or Majorana neutrinos in the final state, can be found in
refs.~\cite{Kim:2021dyj,Kim:2022xjg}.

\section{Parametrization of the SMEFT interactions}
\label{sec:pars}

\subsection{SMEFT input parameter selection}

The standard way of parameterizing the SMEFT Lagrangian is to use the
natural set of couplings defining the dimension-4 renormalizable
interactions (i.e.  the SM Lagrangian) supplied with the Wilson
coefficients of the higher order operators.  The commonly used set of
quantities parameterizing the $d=4$ part of Lagrangian is:
\bea
\bar g,\bar g',\bar g_s & \qquad & SU(2), U(1), SU(3)
\mathrm{~gauge~couplings}\nonumber\\
v, \lambda & \qquad &
\mathrm{Higgs~boson~mass~and~quartic~coupling}\nonumber\\
m_q & \qquad & \mathrm{quark~masses}, q=u,c,t,d,s,b\nonumber\\
K & \qquad & \mathrm{CKM~quark~mixing~matrix}\\
m_\ell, m_{\nu_\ell} & \qquad
& \mathrm{charged~lepton~and~neutrino~masses},
\ell=e,\mu,\tau\nonumber\\
U & \qquad & \mathrm{PMNS~lepton~mixing~matrix}\nonumber
\label{eq:defpar}
\eea
In the list above we assume that gauge couplings $\bar g,\bar g',\bar
g_s$ are already redefined as in eq.~(\ref{Zg-norm}) and $v$ is the
minimum of the full Higgs boson potential, including the higher order
operators.

SMEFT Feynman rules evaluated by \sfr v3 can be expressed in terms of
such set of parameters and WCs of higher dimension operators.  We
further called it to be the ``default'' parametrization set, selected
using {\tt Option $\to$ ``smeft''} in various routines of the code, as
described in Sec.~\ref{sec:install}.  Expressing observables
calculated in SMEFT in terms of ``default'' parameter gives a natural
extension of the corresponding formulae in SM, as the latter can be
immediately obtained by setting all WCs to zero.  However, some
parameters in eq.~(\ref{eq:defpar}), namely gauge and Higgs couplings,
$K$ and $U$ mixing matrices (also particle masses if they are not
chosen to be physical pole masses) are not directly measurable
quantities.  Their numerical values in SMEFT have to be derived by
choosing an appropriate ``input parameter scheme'', i.e.  set of
observables $O_1,\ldots,O_n$ , and expressing them in terms of such
input parameters and WCs:
\bea
\bar g &=& \bar g(O_1,\ldots,O_n,C_i)\;,\nonumber\\
\bar g'&=& \bar g'(O_1,\ldots,O_n,C_i)\;,\nonumber\\
&\ldots \;\;.&
\label{eq:ginpdef}
\eea
Such a procedure leads to additional complications in calculating
processes within SMEFT.  All physical quantities have to be
consistently calculated to a given order of $1/\Lambda$ expansion in
order to keep the result gauge invariant.  Therefore, any observable,
${\cal A}$, calculated in terms of ``default'' parameters of
eq.~(\ref{eq:defpar}) has to be re-expanded to a given EFT order after
expressing in terms of input parameters:
\bea
{\cal A} &=& {\cal A}_4(\bar g,\bar g',\ldots) +
\frac{1}{\Lambda^2}{\cal A}_6^i(\bar g,\bar g',\ldots) C_6^i
\nonumber\\
&+& \frac{1}{\Lambda^4}\left({\cal A}_8^{1ij}(\bar g,\bar g',\ldots)
C_6^i C_6^j + {\cal A}_8^{2i}(\bar g,\bar g',\ldots) C_8^i\right) +
\ldots \nonumber\\
&=& {\cal A'}_4(O_1,O_2,\ldots) + \frac{1}{\Lambda^2}{\cal
  A'}_6^i(O_1,O_2,\ldots) C_6^i \nonumber\\
&+& \frac{1}{\Lambda^4}\left({\cal A'}_8^{1ij}(O_1,O_2,\ldots) C_6^i
C_6^j + {\cal A'}_8^{2i}(O_1,O_2,\ldots) C_8^i\right) + \ldots
\eea
where for simplicity we neglected odd powers in $1/\Lambda$ expansion
as they are always lepton or baryon number violating and strongly
suppressed.

Re-expressing SMEFT amplitudes and re-expanding them in $1/\Lambda$
powers can be technically tedious and error-prone, especially at
higher EFT orders.  Therefore, it is useful to have SMEFT interaction
vertices expressed from the very beginning directly in terms of a set
of measurable physical observables.  Calculations done in terms of
such Feynman rules can be simply truncated at required EFT order,
without the need of re-parametrization.  \sfr v3 provides such
capability of evaluating the SMEFT Lagrangian and interaction vertices
directly in terms of {\em any} user defined set of input parameters.

\subsection{User-defined input parameters}

\sfr v3 allows users to choose their own preferred set of input
parameters, providing they are defined in the correct format and
related to the ``default'' parameters set defined in
eq.~(\ref{eq:defpar}).  The user-defined input parameters in \sfr
should fulfil the following conditions:
\begin{itemize}
\item they are assumed to be measurable physical observables or other
  quantities which do not depend on the SMEFT parameters, in
  particular on WCs of higher dimension operators.
\item they should be real scalar numbers, i.e. do not carry any flavor
  or gauge indices.  If necessary, indexed arrays of flavor or gauge
  parameters should be represented by the relevant set of separate
  scalar entries.
\item names of user-defined parameters should not overlap the names of
  variables already used by the code.  \sfr performs checks for
  overlapping names of variables and displays if necessary relevant
  warnings.
\item user-defined parameters and relations between them and
  ``default'' parameters should be defined in the file {\tt
  code/smeft\_input\_scheme.m}.
\item the format for defining user input parameters follows the
  standard format of \frules model definition files, as illustrated in
  the example below:\\[2mm]
{\tt SM\$InputParameters = \{\\
(* observables used as input parameters in gauge and Higgs sector *)\\
\hspace*{1cm}  alphas == \{\\
\hspace*{2cm}  ParameterType    -> External,\\
\hspace*{2cm}  Value            -> 0.1176,\\
\hspace*{2cm}    InteractionOrder -> \{QCD,2\},\\
\hspace*{2cm}    TeX              -> Subscript[$\backslash$[Alpha],s],\\
\hspace*{2cm}    Description      -> "average alpha\_s at MZ scale"\\
\hspace*{1cm}   \},\\
\hspace*{1cm}$\ldots$\\
\}
}\\[2mm]
A more detailed example of user input parameter definition can be
found in the header of the file {\tt code/smeft\_input\_scheme.m}
supplied with the \sfr v3 distribution.
\item the chosen set of user input parameters must be sufficient to
  fully define ``default'' SMEFT parameters in terms of them and WCs
  of higher dimension operators.  After choosing their own input
  parameters, further referred to as ``input schemes'', the users are
  supposed to provide the corresponding routine with analytical
  expressions for {\em all} variables listed in
  Table~\ref{tab:userinput}.  The example of such a routine and
  predefined most-often used SMEFT input scheme are again provided in
  the file {\tt code/smeft\_input\_scheme.m} (see routine {\tt
    SMEFTInputScheme}).
\end{itemize}

\begin{table}[htb!]
\begin{center}
\begin{tabular}{llllll}
\hline
\multicolumn{2}{c}{Gauge and Higgs sector} & \multicolumn{2}{c}{Quark
  sector} & \multicolumn{2}{c}{Lepton sector} \\
\hline
UserInput\$vev & $v$ & UserInput\$MQU & $m_u$ & UserInput\$MLE &
$m_e$ \\
UserInput\$GW & $\bar g$ & UserInput\$MQC & $m_c$ & UserInput\$MLM &
$m_\mu$ \\
UserInput\$G1 & $\bar g'$ & UserInput\$MQT & $m_t$ & UserInput\$MLT &
$m_\tau$ \\
UserInput\$GS & $\bar g_s$ & UserInput\$MQD & $m_d$ & UserInput\$MVE &
$m_{\nu_e}$ \\
UserInput\$hlambda & $\lambda$ & UserInput\$MQS & $m_s$ &
UserInput\$MVM & $m_{\nu_\mu}$ \\
UserInput\$MZ & $M_Z$ & UserInput\$MQB & $m_b$ & UserInput\$MVT &
$m_{\nu_\tau}$ \\
UserInput\$MW & $M_W$ & UserInput\$CKM & $K$ & UserInput\$PMNS &
$U$ \\ UserInput\$MH & $M_H$ & & & \\
\hline
\end{tabular}
\end{center}
\caption{Names of normalisation constants and corresponding internal
  \sfr variables.  \label{tab:userinput}}
\end{table}

\subsection{Predefined input schemes}
\label{sec:predefs}

Although \sfr v3 in principle allows defining any set of user-defined
input parameters, some input schemes are more natural and technically
easier to use than others.  In particular, it is almost obligatory to
use physical masses of SM particles as part of the input parameter
set.  Otherwise, if masses are calculated as combinations of other
variables and WCs, the latter appear in the particle propagators,
making all amplitude calculations and $1/\Lambda$ expansions
significantly more difficult.  This leaves only $\bar g$, $\bar g'$,
the vev $v$, and $\lambda$ in the electroweak sector, $\bar g_s$ in
the strong sector, CKM matrix $K$ in the quark sector and PMNS matrix
$U$ in the lepton sector to be defined in terms of input parameters.

\sfr v3 provides predefined routines realising the most commonly used
SMEFT input schemes which can be selected by calling the {\tt
  SMEFTInputScheme} routine with relevant options:
\begin{itemize}
\item Gauge sector:
\begin{itemize}
\item $(G_F, M_Z, M_W, M_H)$ input scheme or
\item $(\alpha_{em}, M_Z, M_W, M_H)$ input scheme
\end{itemize}
where $M_Z,M_W,M_H$ are the physical gauge and Higgs boson masses and
$G_F$ is the Fermi constant derived from the muon lifetime.

In both cases ``default'' electroweak sector parameters $\bar g, \bar
g', v$ and $\lambda$ are expressed in terms of input parameters listed
above including linear and quadratic corrections from all contributing
$d=6$ operators and linear corrections from only-bosonic $d=8$
operators.

Strong coupling $\bar g_s$ is defined as $\sqrt{4\pi\alpha_s(M_Z)}$
with some input value of $\alpha_s(M_Z)$ assumed.  Currently, \sfr v3
distribution does not include {\em any} corrections from higher order
operators, leaving it eventually to further modifications by users.
It is not an easy task - the experimental value of $\alpha_s(M_Z)$
cited in literature is an average from various types of measurements.
The correct derivation of such an average in SMEFT should take into
account the fact that different processes used to determine
$\alpha_s(M_Z)$ are affected in different ways by the presence of the
higher dimension operators, thus the relation of the ``averaged''
$\alpha_s(M_Z)$ to $\bar g_s$ has a complicated dependence on WCs of
such operators.  To our knowledge, no such analysis exists yet in the
literature, providing formulae which could be implemented in the
symbolic or numerical codes.

\item Quark sector: Quark masses are assumed to be their physical
  masses - even if such notion is unclear in case of light quarks,
  their values usually do not affect in substantial way most of
  practical calculations, so also the exact definitions are not so
  important in this case.  Corrections to CKM matrix $K$ are evaluated
  using the formulae derived in ref.~\cite{Descotes-Genon:2018foz}.
  They are accurate up to $d=6$ linear terms.

One should note that non-vanishing values of some flavor off-diagonal
4-quark operators can lead to numerically very large corrections to
CKM elements.  If they are larger than 20\%, \sfr v3 displays a
relevant warning and does not include corrections to CKM matrix at
all.  They can be forced to be included independently on how large
they appear using the option {\tt CKMInput $\to$ "force"} in {\tt
  SMEFTInitializeModel} routine.

\item Lepton sector:  Charged lepton masses are assumed to be physical
  masses.  Neutrino masses are calculated as proportional to the WC of
  $d=5$ Weinberg operator, $m_{\nu_i} = v^2 |C_{\nu\nu}^i|$.  The PMNS
  matrix is currently evaluated from measured neutrino mixing angles
  without including corrections from higher order operators, again
  leaving it to eventual future modifications by users.
\end{itemize}

In the predefined input scheme routines in the gauge sector, all
re-parametrizations are done analytically.  Analytical formulae for
corrections to $K$ matrix element are lengthy and complicated, leading
to very long and hardly readable expressions for interaction vertices
and as result also transition amplitudes.  Therefore, currently,
corrections to CKM matrix elements from the $d=6$ operators are in
\sfr v3 evaluated numerically and added to SM central values.

\subsection{Output parametrization}
\label{sec:outpar}

Following the options described above, \sfr v3 can calculate the
interaction vertices in mass basis parametrized in three
(user-selectable) forms:
\begin{enumerate}
\item The ``unexpanded'' (selected as option {\tt Expansion $\to$
  "none"} in relevant routines as described in Sec.~\ref{sec:sfr})
  parametrization.  Interaction vertices are given in terms of
  ``default'' parameters, WCs and $Z_X$ normalisation constants
  without expressing the latter explicitly in terms of ``default'' or
  ``user-defined'' parameters.  Such output is compact and fast to
  produce.  Also, it is the most universal one - adding additional
  higher order operators (like fermionic $d=8$ operators or even
  higher EFT orders), apart from directly appearing new vertices, can
  be easily accommodated by adding new contributions to expressions
  for $Z_X$.  However, in such form, consistent expansion to a given
  EFT order is hidden and can be done only after substituting explicit
  expressions for $Z_X$.  Sample vertices in such parametrization are
  displayed in Fig.~\ref{fig:vert_noexp}.

\begin{figure}
\begin{minipage}{.3\textwidth}
    \centering \epsfig{file=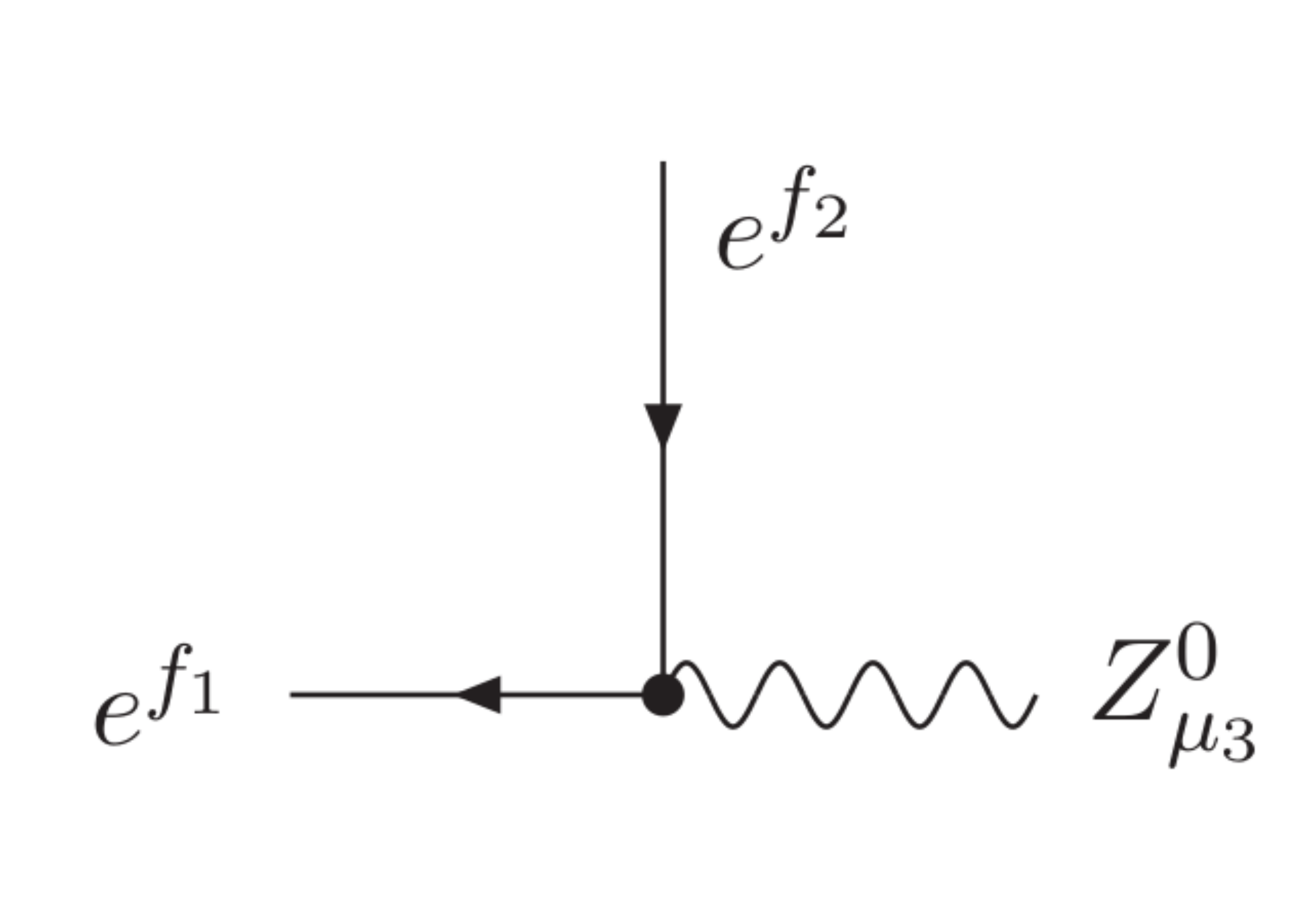,scale=0.17}
  \end{minipage}
  \begin{minipage}{.75\textwidth}
  \footnotesize
   \begin{eqnarray*}
& + & \frac{i}{2}\delta_{f_1 f_2} \left({\bar g}^\prime Z_{g^{\prime}}
     Z_{\gamma_Z}^{21} \left(\gamma^{\mu_3} P_L + 2 \gamma^{\mu_3}
     P_R\right) + {\bar g}{} Z_g Z_{\gamma_Z}^{11} \gamma^{\mu_3}
     P_L\right)\\
& - & \sqrt{2} vZ_{\gamma_Z}^{21} p_3^{\nu} \left(C^{eB*}_{f_2 f_1}
     \sigma^{\mu_3 \nu } P_L + C^{eB}_{f_1 f_2} \sigma^{\mu_3 \nu }
     P_R \right)\\
& + & \sqrt{2} vZ_{\gamma_Z}^{11} p_3^{\nu} \left(C^{eW*}_{f_2 f_1}
     \sigma^{\mu_3 \nu } P_L + C^{eW}_{f_1 f_2} \sigma^{\mu_3 \nu }
     P_R \right)\\
& + & \frac{i v^2}{2} \gamma^{\mu_3} P_R \left({\bar g}{} Z_g
     Z_{\gamma_Z}^{11} - {\bar g}^\prime Z_{g^{\prime}}
     Z_{\gamma_Z}^{21}\right) C^{ \varphi e}_{f_1 f_2}\\
& + & \frac{i v^2}{2} \gamma^{\mu_3} P_L \left({\bar g}{} Z_g
     Z_{\gamma_Z}^{11} - {\bar g}^\prime Z_{g^{\prime}}
     Z_{\gamma_Z}^{21}\right) C^{ \varphi l1}_{f_1 f_2}\\
& + & \frac{i v^2}{2} \gamma^{\mu_3} P_L \left({\bar g}{} Z_g
     Z_{\gamma_Z}^{11} - {\bar g}^\prime Z_{g^{\prime}}
     Z_{\gamma_Z}^{21}\right) C^{ \varphi l3}_{f_1 f_2}
\end{eqnarray*}
\normalsize
  \end{minipage}%
  \newline
\begin{minipage}{.3\textwidth}
    \centering
    \epsfig{file=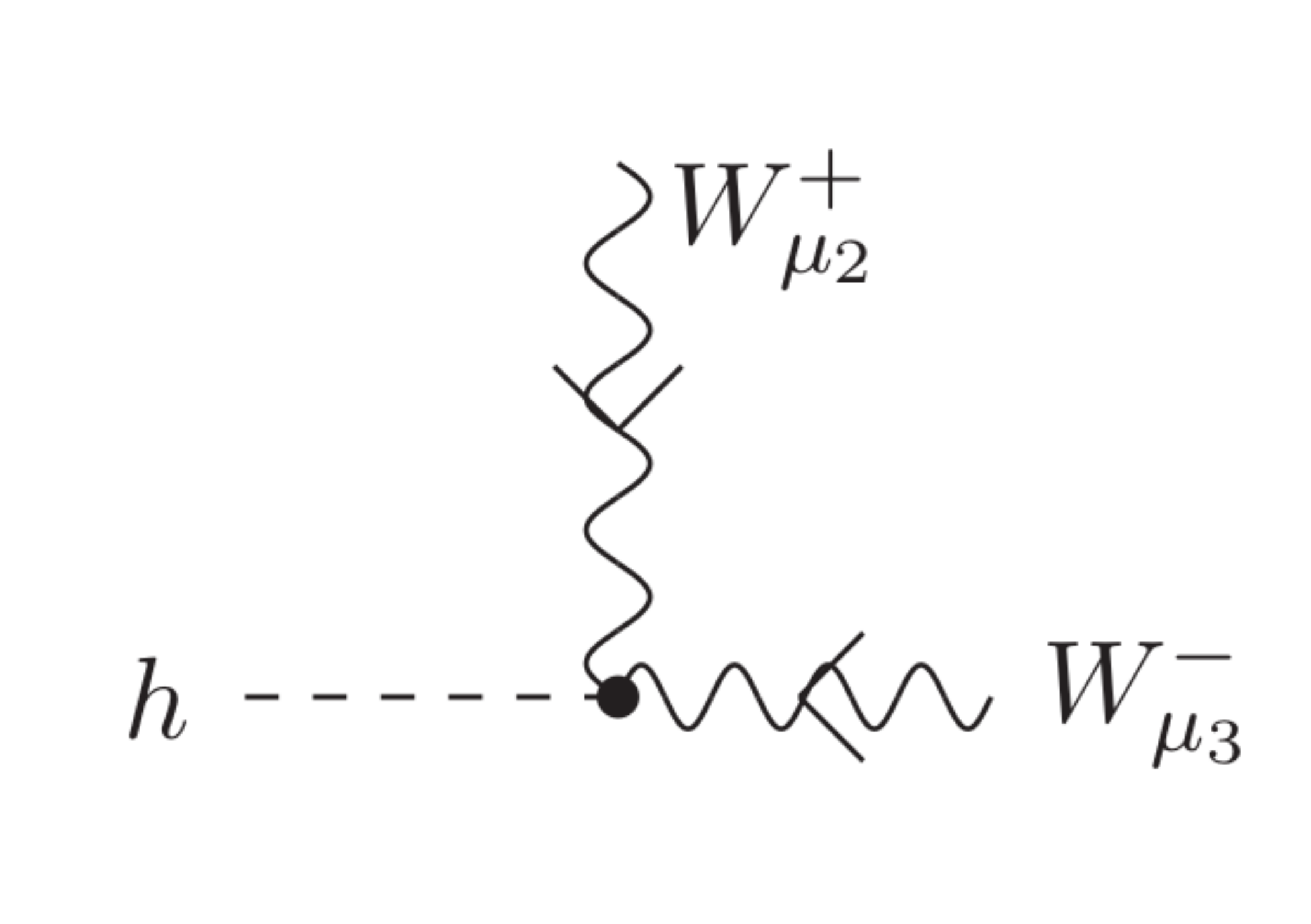,scale=0.17}
  \end{minipage}
  \begin{minipage}{.75\textwidth}
  \footnotesize
\begin{eqnarray*}
  + \frac{i {\bar g}{}^2 v}{2 Z_h}\eta_{\mu_2 \mu_3} + \frac{4 i
    v}{Z_g^2 Z_h}\left(p_2^{\mu_3} p_3^{\mu_2} -
  p_{2}\nobreak\cdot\nobreak{}p_{3} \eta_{\mu_2 \mu_3}\right) C^{
    \varphi W}
\end{eqnarray*}
\normalsize
  \end{minipage}%
  \captionof{figure}{$Z\ell^+\ell^-$ and $hW^+W^-$ vertices before
    expansion of $Z_X$ couplings (including a sample list of operators
    up to maximal dimension-6).  For simplicity in displaying every
    Feynman rule, the $1/\Lambda^2$-factor accompanying every $d=6$
    Wilson Coefficient is omitted e.g.  $C^{\varphi W}\to C^{\varphi
      W}/\Lambda^2$.
\label{fig:vert_noexp}}
\end{figure}

\item The ``default'' (chosen by the option {\tt Expansion $\to$
  "smeft"}) parametrization.  Interaction vertices are given in terms
  of ``default'' parameters and WCs, with shifts of SM fields and
  couplings expanded accordingly.  The result is truncated to
  user-selectable EFT order ($d=4$, $6$ or $8$).  Sample vertices in
  such parametrization are displayed in Fig.~\ref{fig:vert_smeft}.

\begin{figure}[h!]
\begin{minipage}{.3\textwidth}
    \centering
    \epsfig{file=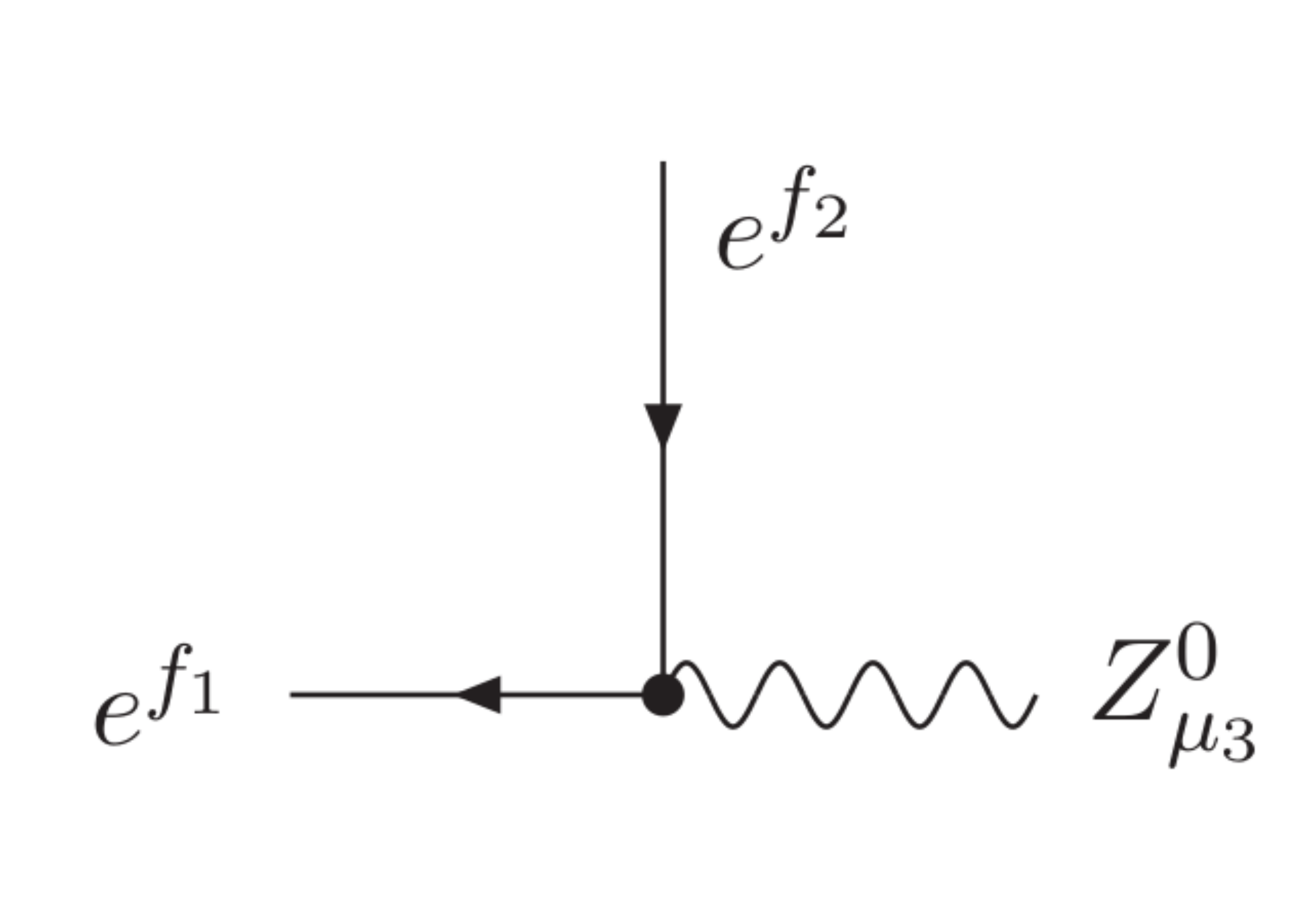,scale=0.17}
  \end{minipage}
  \begin{minipage}{.75\textwidth}
  \footnotesize
\begin{eqnarray*}
&-& \frac{i}{2 \sqrt{{\bar g}^{\prime 2} + {\bar g}{}^2}}\delta_{f_1
    f_2} \left(\left({\bar g}^{\prime 2} - {\bar g}^2\right)
  \gamma^{\mu_3} P_L + 2 {\bar g}^{\prime 2} \gamma^{\mu_3}
  P_R\right)\\
&+& \frac{i {\bar g}^\prime {\bar g}{} v^2}{2 \left({\bar g}^{\prime
      2} + {\bar g}{}^2\right)^{3/2}}\delta_{f_1 f_2}
  \left(\left({\bar g}^{\prime 2} - {\bar g}{}^2\right) \gamma^{\mu_3}
  P_L - 2 {\bar g}{}^2 \gamma^{\mu_3} P_R\right) C^{ \varphi WB} \\
&+& \frac{\sqrt{2} {\bar g}^\prime v}{\sqrt{{\bar g}^{\prime 2} +
      {\bar g}{}^2}}p_3^{\nu} \left(C^{eB*}_{f_2 f_1} \sigma^{\mu_3
    \nu } P_L + C^{eB}_{f_1 f_2} \sigma^{\mu_3 \nu } P_R \right)\\
&+& \frac{\sqrt{2} {\bar g}{} v}{\sqrt{{\bar g}^{\prime 2} + {\bar
        g}{}^2}}p_3^{\nu} \left(C^{eW*}_{f_2 f_1} \sigma^{\mu_3 \nu }
  P_L + C^{eW}_{f_1 f_2} \sigma^{\mu_3 \nu } P_R \right)\\
&+& \frac{1}{2} i v^2 \sqrt{{\bar g}^{\prime 2} + {\bar g}{}^2} C^{
    \varphi e}_{f_1 f_2} \gamma^{\mu_3} P_R + \frac{1}{2} i v^2
  \sqrt{{\bar g}^{\prime 2} + {\bar g}{}^2}C^{ \varphi l1}_{f_1 f_2}
  \gamma^{\mu_3} P_L \\
 & +&  \frac{1}{2} i v^2 \sqrt{{\bar g}^{\prime 2} +
    {\bar g}{}^2}C^{ \varphi l3}_{f_1 f_2} \gamma^{\mu_3} P_L
\end{eqnarray*}
\normalsize
  \end{minipage}%

\begin{minipage}{.3\textwidth}
    \centering
    \epsfig{file=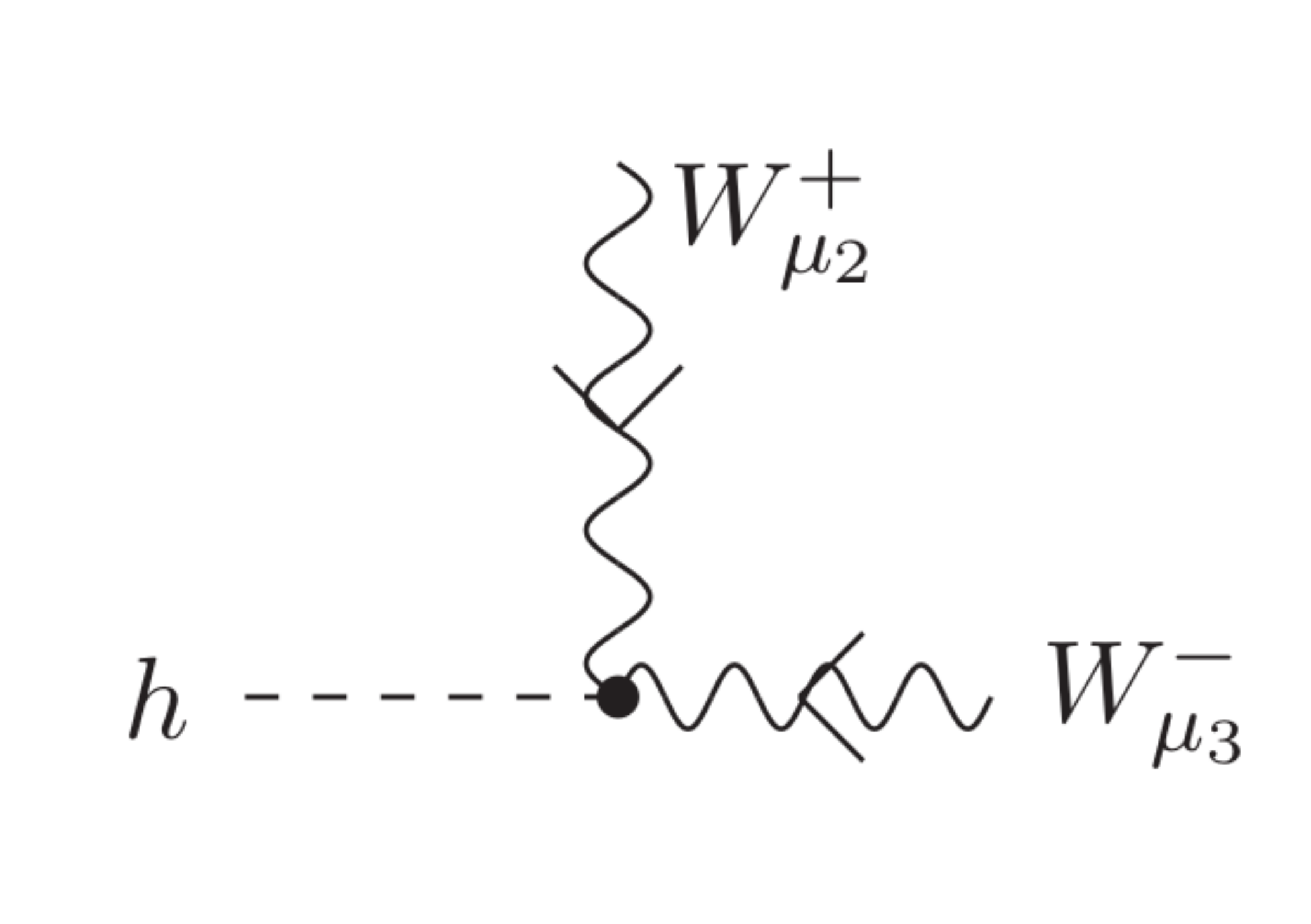,scale=0.17}
  \end{minipage}
  \begin{minipage}{.75\textwidth}
\footnotesize
\begin{eqnarray*}
&+& \frac{1}{2} i {\bar g}{}^2 v\eta_{\mu_2 \mu_3} + \frac{1}{2} i
  {\bar g}{}^2 v^3\eta_{\mu_2 \mu_3} C^{ \varphi \Box} - \frac{1}{8} i
  {\bar g}{}^2 v^3\eta_{\mu_2 \mu_3} C^{ \varphi D}\\
&+& 4 i vC^{ \varphi W} \left(p_2^{\mu_3} p_3^{\mu_2} -
  p_{2}\nobreak\cdot\nobreak{}p_{3} \eta_{\mu_2 \mu_3}\right)
\end{eqnarray*}
\normalsize
  \end{minipage}%
\caption{Same as in Fig.~\ref{fig:vert_noexp} but in default ($\bar
  g'$, $\bar g$, $v$) parametrization scheme (the $Z_X$ couplings are
  expanded up to maximal dimension-6 terms).
\label{fig:vert_smeft}}
\end{figure}

\item The ``user'' (chosen by the option {\tt Expansion $\to$ "user"})
  parametrization.  Interaction vertices are given directly in terms
  of user-defined input parameters and WCs, again with shifts of SM
  fields and couplings expanded accordingly.  The result is truncated
  to user-selectable EFT order ($d=4$, $6$ or $8$).  Sample vertices
  for the $(G_F,M_Z,M_W,M_h)$ input scheme in the electroweak sector
  (see discussion in Sec.~\ref{sec:predefs}) are displayed in
  Fig.~\ref{fig:vert_gf}.

\begin{figure}[h]
\begin{minipage}{.3\textwidth}
    \centering
    \epsfig{file=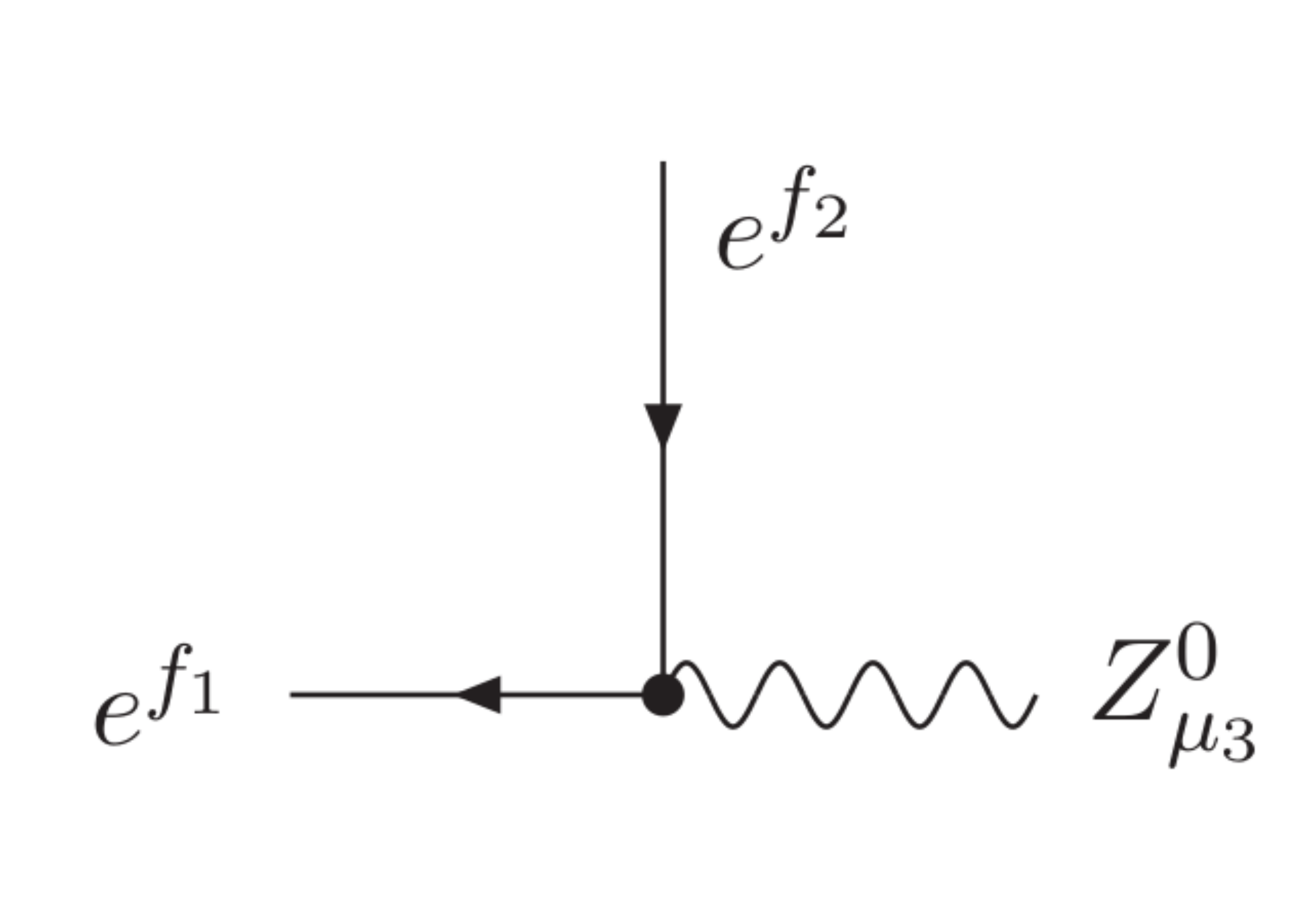,scale=0.17}
  \end{minipage}
  \begin{minipage}{.75\textwidth}
\footnotesize
\begin{eqnarray*}
&-& \frac{i 2^{1/4}\sqrt{G_F}}{M_Z}\delta_{f_1
    f_2}\left(\left(M_Z^2-2M_W^2\right)\gamma^{\mu_3} P_L + 2
  \left(M_Z^2-M_W^2\right) \gamma^{\mu_3} P_R\right)\\
&+& \frac{i 2^{3/4} M_W \sqrt{M_Z^2-M_W^2}}{M_Z\sqrt{G_F}} \delta_{f_1
    f_2} C^{ \varphi WB} \gamma^{\mu_3}\\
&+& \frac{2^{1/4}\sqrt{M_Z^2-M_W^2}}{\sqrt{G_F}
   M_Z} p_3^{\nu} \left(C^{eB*}_{f_2 f_1} \sigma^{\mu_3 \nu } P_L +
  C^{eB}_{f_1 f_2} \sigma^{\mu_3 \nu } P_R \right) \\
&+& \frac{2^{1/4} M_W}{\sqrt{G_F} M_Z} p_3^{\nu} \left(C^{eW*}_{f_2
    f_1} \sigma^{\mu_3 \nu } P_L + C^{eW}_{f_1 f_2} \sigma^{\mu_3 \nu
  } P_R \right)\\
&+& \frac{i\, \delta_{f_1 f_2}}{2^{9/4}\sqrt{G_F} M_Z} C^{ \varphi D}
  \left(\left(2M_W^2+M_Z^2\right) \gamma^{\mu_3} P_L + 2
  \left(M_W^2+M_Z^2\right) \gamma^{\mu_3} P_R\right) \\
&+&\frac{i M_Z}{2^{1/4} \sqrt{G_F}}C^{ \varphi e}_{f_1 f_2}
  \gamma^{\mu_3} P_R + \frac{i M_Z}{2^{1/4} \sqrt{G_F}}C^{ \varphi
    l1}_{f_1 f_2} \gamma^{\mu_3} P_L + \frac{i M_Z}{2^{1/4}
    \sqrt{G_F}} C^{ \varphi l3}_{f_1 f_2} \gamma^{\mu_3} P_L
  \\
&+& \frac{i\,\delta_{f_1 f_2}}{2^{9/4}\sqrt{G_F}
    M_Z}C^{ll}_{2112}\left(\left(M_Z^2-2M_W^2\right)\gamma^{\mu_3} P_L
  + 2 \left(M_Z^2-M_W^2\right) \gamma^{\mu_3} P_R\right) \\
&+&\frac{i\,\delta_{f_1 f_2}}{2^{9/4}\sqrt{G_F} M_Z}\left(C^{\varphi
    l3}_{11}+C^{\varphi
    l3}_{22}\right)\left(\left(2M_W^2-M_Z^2\right)\gamma^{\mu_3} P_L +
  2 \left(M_W^2-M_Z^2\right)\gamma^{\mu_3} P_R\right)
\end{eqnarray*}
\normalsize
  \end{minipage}%
\newline
\begin{minipage}{.3\textwidth}
    \centering
    \epsfig{file=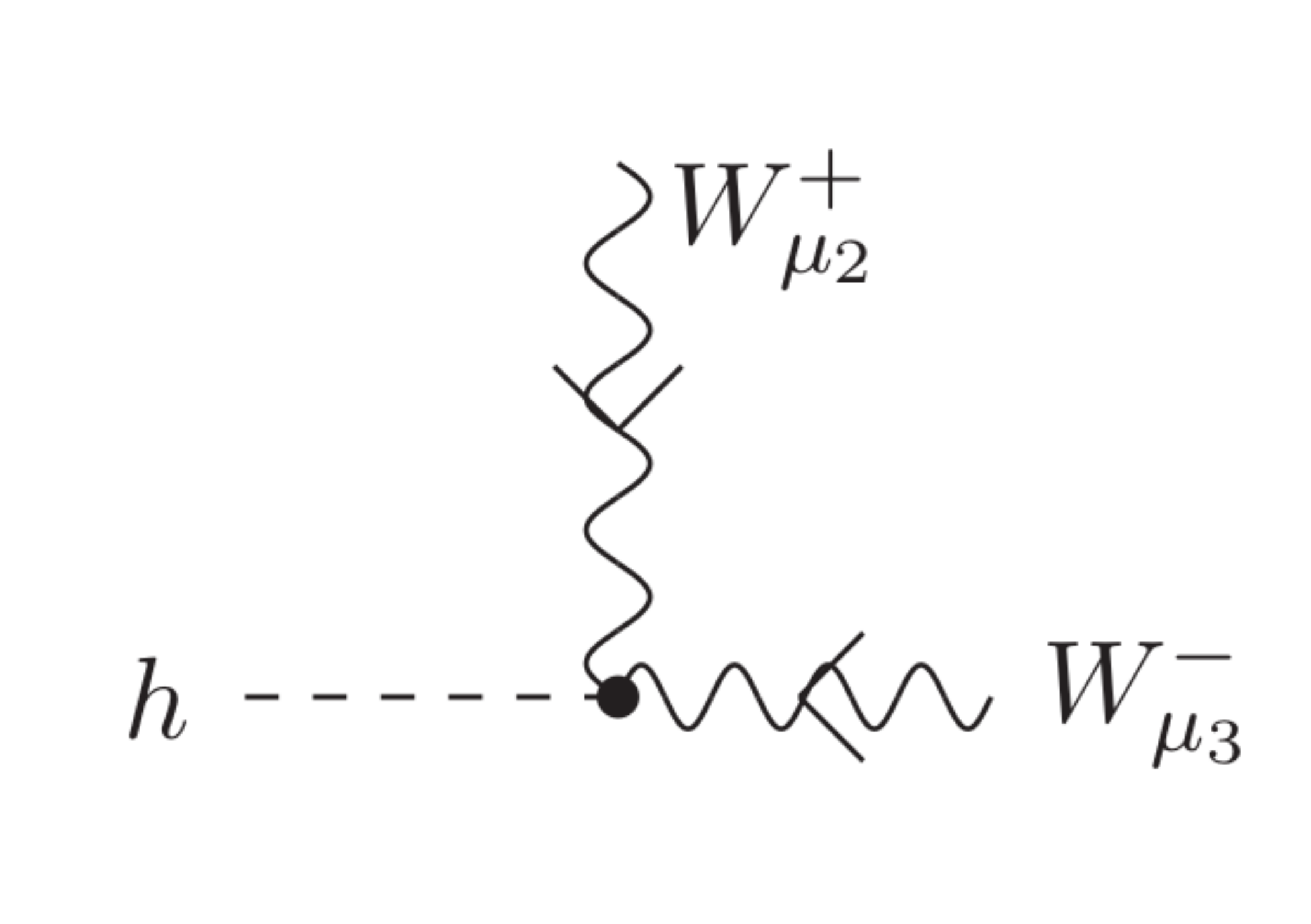,scale=0.17}
  \end{minipage}
  \begin{minipage}{.75\textwidth}
\footnotesize
\begin{eqnarray*}
& +& i 2^{3/4} \sqrt{G_F} M_W^2 \eta_{\mu_2 \mu_3} +\frac{i 2^{3/4}
   M_W^2}{\sqrt{G_F}} \eta_{\mu_2 \mu_3} C^{ \varphi \Box} - \frac{i
    M_W^2}{2^{3/4}\sqrt{G_F}} \eta_{\mu_2 \mu_3} C^{ \varphi D}\\
& -& \frac{i M_W^2}{2^{3/4}\sqrt{G_F}} \eta_{\mu_2 \mu_3}
  C^{ll}_{2112} + \frac{i M_W^2}{2^{3/4}\sqrt{G_F}} \eta_{\mu_2 \mu_3}
  \left( C^{\varphi l 3}_{11} + C^{\varphi l 3}_{22}\right) \\
&+& \frac{ i 2^{7/4}}{\sqrt{G_F}} C^{ \varphi W} \left(p_2^{\mu_3}
  p_3^{\mu_2} - p_{2}\nobreak\cdot\nobreak{}p_{3} \eta_{\mu_2
    \mu_3}\right)
%
% AEM scheme:  
%+ \frac{2 i \sqrt{\alpha_{em}} \sqrt{\pi} M_W M_Z \eta_{\mu_2 \mu_3}
%}{\sqrt{M_Z^2-M_W^2}} + \frac{2 i M_W^4 \eta_{\mu_2 \mu_3} C^{\varphi
%WB}}{ \sqrt{\alpha_{em}} \sqrt{\pi} M_Z } +\frac{2 i (M_W^3 M_Z^2 -
%M_W^5) \eta_{\mu_2 \mu_3} C^{ \varphi \Box} }{ \sqrt{\alpha_{em}}
%\sqrt{\pi} M_Z \sqrt{M_Z^2-M_W^2}} - \frac{i(M_W^3 M_Z^2 -
%2M_W^5)\eta_{\mu_2 \mu_3} C^{ \varphi D} }{2 \sqrt{\alpha_{em}}
%\sqrt{\pi} M_Z \sqrt{M_Z^2-M_W^2}} + \frac{4 i M_W \sqrt{M_Z^2-M_W^2}
%C^{ \varphi W} \left(p_2^{\mu_3} p_3^{\mu_2} -
%p_{2}\nobreak\cdot\nobreak{}p_{3} \eta_{\mu_2 \mu_3}\right)}{
%\sqrt{\alpha_{em}} \sqrt{\pi} M_Z }
\end{eqnarray*}
\normalsize
  \end{minipage}%
  \captionof{figure}{Same as in Fig.~\ref{fig:vert_noexp} but in the
    $(G_F,M_Z,M_W,M_h)$ input scheme (the $Z_X$ couplings are expanded
    up to maximal dimension-6 terms).
\label{fig:vert_gf}}
\end{figure}

\end{enumerate}

As described in more details in the next Section, the form of the
output can be selected by choosing various code options.

\section{\sfr installation and code structure}
\label{sec:install}

\subsection{Installation}

\sfr package works using the \frules system, so both need to be
properly installed first.  A recent version and installation
instructions for the \frules package can be downloaded from the
address:
\begin{center}
\url{https://feynrules.irmp.ucl.ac.be}
\end{center}
\sfr v3 has been tested with \frules version 2.3.49.  It should be
used with {\it Mathematica} version 12.1 or later, as also the newest
\frules version was modified to be compatible with {\it Mathematica}
upgrades.

Standard \frules installation assumes that the new models' description
is put into {\tt Models} sub-directory of its main tree.  We follow
this convention, so that the \sfr file archive should be unpacked into
\begin{center}
 {\tt Models/SMEFT\_N\_NN }
\end{center}
catalogue, where {\tt N\_NN} denotes the package version (currently
version {\tt 3\_00}).  After installation, {\tt Models/SMEFT\_N\_NN }
contains the following files and sub-directories listed in
Table~\ref{tab:filestruct}.
  
\begin{table}[htb!]
\begin{mdframed}[backgroundcolor=lightgray,userdefinedwidth=\textwidth]
\begin{tabular}{p{43mm}p{100mm}}
{{\tt SmeftFR-init.nb smeft\_fr\_init.m }} & Notebook and equivalent
text script generating SMEFT Lagrangian in mass basis and Feynman
rules in {\em Mathematica} format.  \\[2mm]
{{\tt SmeftFR-interfaces.nb smeft\_fr\_interfaces.m}} & Notebook and
text script with routines for exporting Feynman rules in various
formats: WCxf, Latex, UFO and FeynArts.\\[2mm]
{\tt SmeftFR\_v3.pdf} & package manual in pdf format.\\[2mm]
{\tt code} & sub-directory with package code and utilities.\\[2mm]
{\tt lagrangian} & sub-directory with expressions for the SM
Lagrangian and dimension-5, 6 and 8 operators coded in \frules
format.\\[2mm]
{\tt definitions} & sub-directory with templates of SMEFT ``model
files'' and example of numerical input for Wilson coefficients in WCxf
format.\\[2mm]
{\tt output} & sub-directory with dynamically generated model
``parameter files'' and output for Feynman rules in various formats,
by default {\em Mathematica}, Latex, UFO and FeynArts are
generated.\\[2mm]
\end{tabular}
\end{mdframed}
\caption{Files and directories included in \sfr v3.00
  package.  \label{tab:filestruct} }
\end{table}

Before running the package, one needs to set properly the main \frules
installation directory, defining the {\tt \$FeynRulesPath} variable at
the beginning of {\tt smeft\_fr\_init.m} and {\tt
  smeft\_fr\_interfaces.m} files.  For non-standard installations (not
advised!), also the variable {\tt SMEFT\$Path} has to be updated
accordingly.

\subsection{Code structure}
\label{sec:code}

The most general version of SMEFT, including all possible flavour
violating couplings, is very complicated.  Symbolic operations on the
full SMEFT Lagrangian, including the complete set of dimension-5 and-6
operators and bosonic dimension-8 operators, with numerical values of
all Wilson coefficients assigned, are time-consuming and can take
hours or even days on a standard personal computer.  For most of the
physical applications it is sufficient to derive interactions only for
a subset of operators.\footnote{Eventually, operators must be selected
with care as in general they may mix under
renormalisation~\cite{Jenkins:2013zja, Jenkins:2013wua,
  Alonso:2013hga}.}

\begin{figure}[htb!]
\includegraphics[width=\textwidth,height=0.91\textheight]{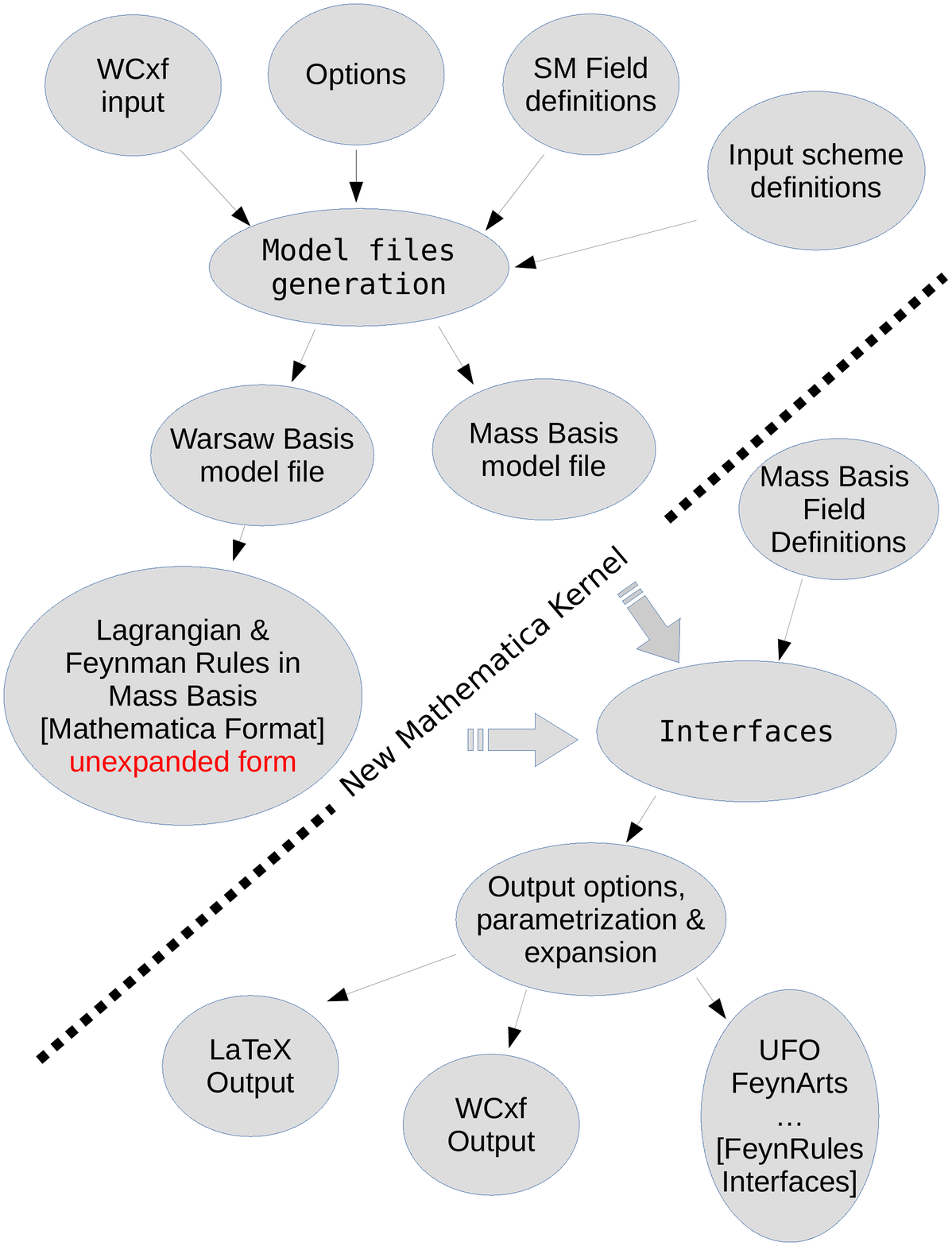}
  \caption{Structure of the \sfr v3 code.   \label{fig:flow}}
\end{figure}

To speed up the calculations, \sfr can evaluate Feynman rules for a
chosen subset of operators only, generating dynamically the proper
\frules ``model files''.  The calculations are divided in three
stages, as illustrated in the flowchart of Fig.~\ref{fig:flow}.
\begin{itemize}
\item First, before initialising the \frules engine, a routine
  relating default and user-defined input parameters are executed.
  Numerical values of parameters depending on WCs of higher order
  operators are calculated.  Then, two \frules model files for SMEFT
  (for gauge and mass basis) are dynamically generated, containing all
  variables required to fully describe interactions in various
  parametrizations (see Sec.~\ref{sec:outpar}).
\item Next, the SMEFT Lagrangian is initialised in gauge basis and
  transformed to mass eigenstates basis analytically.  At this stage,
  $Z_X$ normalisation constants are evaluated in terms of both
  ``default'' and ``user-defined'' input parameters, but such explicit
  expressions are not substituted in interaction vertices.  This very
  significantly speeds up the calculations (approximately by an order
  of magnitude comparing to \sfr v2) and produces expressions that are
  remarkably compact for such a complicated model.  All terms which
  are explicitly of order in $1/\Lambda$ higher than requested by
  users (maximum $1/\Lambda^4$) are truncated, but for consistent
  $1/\Lambda$ expansions such terms must be neglected once more after
  inserting an explicit expression for $Z_X$.  The resulting mass
  basis Lagrangian, normalisation constants $Z_X$ and Feynman rules
  written in Mathematica format are stored on disk.
\item Finally, the previously generated output can be used to export
  mass basis SMEFT interactions in various commonly used external
  formats such as Latex, WCxf and standard \frules supported
  interfaces -- UFO, FeynArts and others.  At this stage, users can
  choose the form of output parametrization, with $Z_X$ normalisation
  constants also replaced by their corresponding explicit forms.
\end{itemize}

\section{Deriving SMEFT Feynman rules  with \sfr package}
\label{sec:sfr}

\subsection{Model initialisation}
\label{sec:init}

\begin{table}[htb!]
\begin{mdframed}[backgroundcolor=lightgray, userdefinedwidth=\textwidth,
      rightmargin=-10mm]
\begin{tabular}{lp{27mm}p{95mm}}
Option & Allowed values & Description \\[2mm]
\hline
\\
Operators & list of operators & Subset of SMEFT operators included in
calculations.  Default: all $d=5$ and $d=6$ operators.  \\[3mm]
Gauge & {\bf Unitary}, Rxi & Choice of gauge fixing conditions.
\\[3mm]
ExpansionOrder & 0, {\bf 1} or 2 & SMEFT interactions are expanded to
$1/\Lambda^{2 \, \mathrm{ExpansionOrder}}$ (default:
$1/\Lambda^{2}$).\\[3mm]
WCXFInitFile & {\bf ""} & Name of file with numerical values of Wilson
coefficients in the WCxf format.  If this option is not set, all WCs
are initialised to $0$.\\[3mm]
RealParameters & {\bf False}, True & Some codes like MadGraph 5 accept
only real values of parameters.  If this option is set to True,
imaginary part of complex parameters
%($K$ and $U$ matrices, WCs)
are truncated in \frules model files.\\[3mm]
InputScheme & {\bf "GF"}, "AEM", \dots & Selection of input parameters
scheme, see discussion in Sec.~\ref{sec:pars}.  \\[3mm]
CKMInput & "no", {\bf "yes"}, "force" & Decides if corrections to CKM
matrix are included (use "force" to add them even their relative size
exceeds the threshold defined in variable {\tt SMEFT\$CKMTreshold}
(default: $0.2$).  \\[3mm]
MaxParticles & {\bf 6} & Only Feynman rules with less then
MaxParticles external legs are calculated.  Does not affect UFO and
FeynArts output.\\[3mm]
MajoranaNeutrino & {\bf False}, True & Neutrinos are treated as
Majorana spinors if $Q_{\nu\nu}$ is included in the operator list or
this option is set to True, massless Weyl spinors otherwise.  \\[3mm]
Correct4Fermion & False, {\bf True} & Corrects relative sign of some
4-fermion interactions, fixing results of \frules.  \\[3mm]
WBFirstLetter & {\bf "c"} & Customisable first letter of Wilson
coefficient names in Warsaw basis (default $c_G, \ldots$).  \\[3mm]
% (to avoid naming clashes with other SMEFT bases).
%
MBFirstLetter & {\bf "C"} & Customisable first letter of Wilson
coefficient names in mass basis (default $C_G, \ldots$).
\end{tabular}
\end{mdframed}  
\caption{The allowed options of {\tt SMEFTInitializeModel} routine.
  If an option is not specified, the default value (marked above in
  boldface) is assumed.
\label{tab:init}
}
\end{table}

In the first step, the relevant \frules model files must be generated.
This is done by calling the function:

\medskip

{\tt SMEFTInitializeModel}[{\it Option1 $\to$ Value1, Option2 $\to$
    Value2, $\ldots$}]

\medskip

\noindent with the allowed options listed in Table~\ref{tab:init}.

The list and the naming of operators employed by \sfr v3 is arranged
and explained in \ref{app:ops}.  By default, all possible 59+1+4 SMEFT
($d=5$ and $d=6$) operator classes and no $d=8$ operators are included
in calculations, the latter can be added trivially by users if
necessary.

To speed up the derivation of Feynman rules and to get more compact
expressions, the user can restrict the list above to any preferred
subset of operators (an example of initialisation with a sample
operator subset is given in Sec.~\ref{sec:sample}).

\sfr is fully integrated with the WCxf standard.  Apart from
numerically editing Wilson coefficients in \frules model files,
reading them from the WCxf input is the only way of automatic
initialisation of their numerical values.  Such an input format is
exchangeable between a larger set of SMEFT-related public
packages~\cite{Aebischer:2017ugx} and helps in comparing their
results.

An additional advantage of using WCxf input format comes in the
flavour sector of the theory.  Here, Wilson coefficients are in
general tensors with flavour indices, in many cases symmetric under
various permutations.  WCxf input requires initialisation of only the
minimal set of flavour dependent Wilson coefficients, those which
could be derived by permutations are also automatically properly
set.\footnote{We would like to thank D.~Straub for supplying us with a
code for symmetrization of flavour-dependent Wilson coefficients.}

There is no commonly accepted standard for initialisation of numerical
values of WCs of $d=8$ operators, but as we only including scalar (no
flavor indices) bosonic operators, adding them to WCxf-type input
files is straightforward, we follow the convention for $d=6$ bosonic
operators just using the new names for $d=8$ entries.

Further comments concern {\tt MajoranaNeutrino} and {\tt
  Correct4Fermion} options.  They are used to modify the analytical
expressions only for the Feynman rules, not at the level of the mass
basis Lagrangian from which the rules are derived.  This is because
some \frules interfaces, like UFO, intentionally leave the relative
sign of 4-fermion interactions uncorrected\footnote{B.~Fuks, private
communication.}, as it is later changed by Monte-Carlo generators like
MadGraph5.  Correcting the sign before generating UFO output would
therefore lead to wrong final result.  Similarly, treatment of
neutrinos as Majorana fields could not be compatible with hard coded
quantum number definitions in various packages.  On the other hand, in
the manual or symbolic computations it is convenient to have from the
start the correct form of Feynman rules, as done by \sfr when both
options are set to their default values.

Currently, the predefined input scheme for initialisation of CKM
matrix elements is based on the approach of
ref.~\cite{Descotes-Genon:2018foz}.  It can lead to numerically very
large corrections to CKM matrix from some of the flavor off-diagonal
4-quark dimension-6 operators.  Such large corrections usually mean
that the assumed values of 4-quark WCs violate experimental bounds on
flavor transitions and should be modified.  In such case, by default
\sfr v3 displays a relevant warning and does not include corrections
to CKM matrix at all, expecting WC values to be modified.  Such
behaviour can be overwritten (so that even huge corrections are
included, but the warning is still displayed) setting option {\tt
  ForceCKMInput $\to$ True}.  The maximum allowed size of corrections
to CKM any of CKM elements is defined by variable {\tt
  SMEFT\$CKMTreshold} in the file {\tt code/smeft\_variables.m} and by
default set to {\tt SMEFT\$CKMTreshold=0.2}.  Users can modify this
number to their preferred sensitivity level.

After execution, {\tt SMEFTInitializeModel} creates in the {\tt
  output} sub-directory two model files:
\begin{itemize}
\item
{\tt smeft\_par\_WB.fr}: SMEFT parameter file with Wilson coefficients
in gauge basis (defined as ``Internal'', with no numerical values
assigned).
\item
{\tt smeft\_par\_MB.fr}: SMEFT parameter file with Wilson coefficients
in mass basis (defined as ``External'', numerical values of WCs
imported from the input file in WCxf format).
\end{itemize}
Note that field definitions are not generated dynamically and stored
as fixed files named {\tt smeft\_fields\_WB.fr} and {\tt
  smeft\_fields\_MB.fr} in {\tt definitions} sub-directory.

Parameter files generated by {\tt SMEFTInitializeModel} contain also
definitions of SM parameters, copied from several template files in
{\tt definitions} sub-directory
%{\tt smeft\_par\_SM.fr}, {\tt smeft\_par\_head\_WB.fr}, {\tt
%smeft\_par\_head\_MB.fr}
and, most importantly, from the header of the {\tt
  code/smeft\_input\_scheme.m} file, where the user-defined input
parameters should be listed.  Only the latter, values of user-defined
parameters, are copied unchanged to model files, numerical values of
other parameters can be updated to include corrections from higher
order operators (thus hand-made modifications in files in {\tt
  definitions} sub-directory are not advised and will be overwritten by
the code).

As mentioned above, in all analytical calculations performed by \sfr,
terms suppressed by terms of the order higher than ${\cal
  O}(1/\Lambda^{2\,\mathrm{ExpansionOrder}})$ are always neglected.
Therefore, the resulting Feynman rules can be consistently used to
calculate physical observables, symbolically or numerically by
Monte-Carlo generators, up to the maximum quadratic order in
dimension-6 operators and linear order in dimension-8 operators.  This
information is encoded in \frules SMEFT model files by assigning the
``interaction order'' parameter to Wilson coefficients: {\tt NP=1} for
$d=6$ WCs and {\tt NP=2} for $d=8$ operators.  {\tt ExpansionOrder}
parameter is passed also to model files {\tt smeft\_par\_WB.fr} and
{\tt smeft\_par\_MB.fr} as:

\bigskip
\begin{tabular}{l}
M\$InteractionOrderLimit = \{ \\
~~~ \{QCD,99\}, \\
~~~ \{NP,ExpansionOrder\}, \\
~~~ \{QED,99\} \\
~ \} 
\end{tabular}
\bigskip

\subsection{Calculation of mass basis Lagrangian and Feynman rules}
\label{sec:mbasis}

By loading the \frules model files, the derivation of SMEFT Lagrangian
in mass basis is performed by calling the following sequence of
routines:
\begin{center}
%
%\begin{mdframed}[backgroundcolor=lightgray]  
%
\begin{tabular}{lp{116mm}}
{\tt SMEFTLoadModel[ ]} & Loads {\tt output/smeft\_par\_WB.par} model
file and imports SMEFT Lagrangian in gauge basis for chosen subset of
operators. \\[2mm]
{\tt SMEFTFindMassBasis[ ]} & Finds field bilinears and analytical
transformations diagonalizing mass matrices.  Calculates the
expressions for $Z_X$ normalisation constants.\\[2mm]
{\tt SMEFTFeynmanRules[ ]} & Evaluates analytically SMEFT Lagrangian
and Feynman rules in the mass basis to a required order in ${\cal
  O}(1/\Lambda)$, {\em without substituting explicit expressions} for
$Z_X$ constants (see example in Fig.~\ref{fig:vert_noexp}).\\[2mm]
{\tt SMEFTOutput[ {\it Options} ]} & By default stores SMEFT model
file with parameters in mass basis as {\tt output/smeft\_par\_MB.m}
and mass basis Lagrangian and vertices in {\tt
  output/smeft\_feynman\_rules.m}.  To generate output in different
locations, use options {\it ModelFile $\to$ filename1} and {\it
  TargetFile $\to$ filename2}.
\end{tabular}
%
%\end{mdframed}
%
\end{center}

The calculation time may vary considerably depending on the choice of
operator (sub-)set and gauge fixing conditions chosen.  For the full
list of SMEFT $d=5$ and $d=6$ operators and in $R_\xi$-gauges, one can
expect CPU time necessary to evaluate all Feynman rules for up to
about an hour on a typical personal computer, depending on its speed
capabilities.  Adding $d=8$ operators can obviously increase the CPU
time, therefore it is advisable to choose only the operators relevant
to a given analysis.

One should note that when neutrinos are treated as Majorana particles,
(as necessary in case of non-vanishing Wilson coefficient of $d=5$
Weinberg operator), their interactions involve lepton number
non-conservation.  Baryon and lepton (BL) number is also not conserved
when explicitly BL-violating 4-fermion operators are included in
Lagrangian. When \frules is dealing with such cases, it produces
warnings of the form:

\bigskip
\noindent \red{{\it QN::NonConserv: Warning: non quantum number
    conserving vertex encountered!}\\[1mm]
{\it Quantum number LeptonNumber not conserved in vertex $\ldots$} }
\bigskip

\noindent Obviously, such warnings in this specific case should be
ignored.

Evaluation of Feynman rules for vertices involving more than two
fermions is not fully implemented yet in {\tt FeynRules}, and some
warnings are displayed.  To our experience, in most cases 4-fermion
vertices are calculated correctly in spite of such warnings, apart
from the issue of relative sign of four fermion diagrams mentioned
earlier.  Some cases are still problematic, e.g.  the correct
automatic derivation of quartic interactions with four Majorana
neutrinos.  For these special cases, \sfr overwrites the \frules
result with manually calculated formulae encoded in Mathematica
format.

Another remark concerns the hermicity property of the SMEFT
Lagrangian.  For some types of interactions, e.g.  four-fermion
vertices involving two-quarks and two-leptons, the function {\tt
  CheckHermicity} provided by \frules reports non-Hermitian terms in
the Lagrangian.  However, such terms are actually Hermitian if
permutation symmetries of indices of relevant Wilson coefficients are
taken into account.  Such symmetries are automatically imposed if
numerical values of Wilson coefficients are initialised with the use
of {\tt SMEFTInitializeMB} or {\tt SMEFTToWCXF} routines (see
Sec.~\ref{sec:interface} and~\ref{sec:wcxf}).

Results of the calculations are by default collected in file {\tt
  output/smeft\_feynman\_rules.m}.  The Feynman rules and parts of the
mass basis Lagrangian for various classes of interactions are stored
in the variables with self-explanatory names listed in
Table~\ref{tab:variables}.

\begin{table}[tb!]
\begin{center}
\begin{mdframed}[backgroundcolor=lightgray]  
{\tt 
\begin{tabular}{lp{20mm}l}
  LeptonGaugeVertices &&   QuarkGaugeVertices \\
  LeptonHiggsGaugeVertices &&  QuarkHiggsGaugeVertices \\
  QuarkGluonVertices \\
  GaugeSelfVertices &&  GaugeHiggsVertices \\
  GluonSelfVertices &&  GluonHiggsVertices \\
  GhostVertices \\
  FourLeptonVertices &&   FourQuarkVertices \\
  TwoQuarkTwoLeptonVertices \\
  DeltaLTwoVertices && BLViolatingVertices 
\end{tabular}
}
\end{mdframed}
\end{center}
\caption{Names of variables defined in the file {\tt
    output/smeft\_feynman\_rules.m} containing expressions for Feynman
  rules.  Parts of mass basis Lagrangian are stored in equivalent set
  of variables, with ``{\tt Vertices}'' replaced by ``{\tt
    Lagrangian}'' in part of their names (i.e.  {\tt
    LeptonGaugeVertices} $\to$ {\tt LeptonGaugeLagrangian},
  \textit{etc.}).  \label{tab:variables}}
\end{table}

File {\tt output/smeft\_feynman\_rules.m} contains also expressions
for the normalisation factors $Z_X$ relating Higgs and gauge fields
and couplings in the Warsaw and mass basis, in ``default'' and
``user'' parametrizations (see Table~\ref{tab:zx} for corresponding
names of code variables).  In addition, formulae for tree level
corrections to SM mass parameters and Yukawa couplings are stored in
variables {\tt SMEFT\$vev}, {\tt SMEFT\$MH}, {\tt SMEFT\$MW}, {\tt
  SMEFT\$MZ}, {\tt SMEFT\$YL[i,j]}, {\tt SMEFT\$YD[i,j]} and {\tt
  SMEFT\$YU[i,j]}, as well as the selected user-defined program
options.

As mentioned before, in expressions for Lagrangian parts and vertices
stored in variables of Table~\ref{tab:variables} the $Z_X$ constants
are left in an unexpanded form, as in Fig.~\ref{fig:vert_noexp}.  To
produce formulae fully expanded in $1/\Lambda$ powers to a required
order, one must call the routine {\tt SMEFTExpandVertices}, e.g.  for
vertices in ``default'' parametrization up to $1/\Lambda^4$ terms one
should use\\[2mm]
{\tt SMEFTExpandVertices[Input -> "smeft", ExpOrder -> 2]}\\[2mm]
(another possible choice is {\tt Input $\to$ "user"}).  Then expanded
version of vertices is copied to variables ending with ``Exp'' ({\tt
  LeptonGaugeVerticesExp, QuarkGaugeVerticesExp} etc.) and can be
displayed or used in further calculations using standard \frules
format.

At this point the Feynman rules for the mass basis Lagrangian are
already calculated, but the definitions for fields and parameters used
to initialise the SMEFT model in \frules are still given in gauge
basis.  To avoid inconsistencies, before exporting calculated
expressions to other formats supported by \frules and \sfr one should
quit the current Mathematica kernel and start a new one reloading the
mass basis Lagrangian together with the compatible model files with
fields defined also in mass basis, as described next in
Sec.~\ref{sec:interface}.  All further calculations should be
performed within this new kernel (routine {\tt SMEFTExpandVertices}
can be also used with this new kernel in the same way as described
above).

\subsection{Output formats and interfaces}
\label{sec:interface}

\sfr output in some of the portable formats must be generated from the
SMEFT Lagrangian transformed to mass basis, with all numerical values
of parameters initialised.  As \frules does not allow for two
different model files loaded within a single \textit{Mathematica}
session, one needs to quit the kernel used to run routines necessary
to obtain Feynman rules and, as described in the previous Section,
start a new \textit{Mathematica} kernel.  Within it, the user must
reload \frules and \sfr packages and call the following routine:

\bigskip

{\tt SMEFTInitializeMB[ {\it Options} ]}

\bigskip

\noindent Allowed options are given in Table~\ref{tab:mbinit}.  After
a call to {\tt SMEFTInitializeMB}, mass basis model files are read and
the mass basis Lagrangian is stored in a global variable named {\tt
  SMEFT\$MBLagrangian} for further use by interface routines.

\begin{table}[htb!]
\begin{center}
\begin{mdframed}[backgroundcolor=lightgray]  
\noindent \begin{tabular}{lp{30mm}p{80mm}}
Option  & Allowed values & Description \\[2mm]
\hline
\\
Expansion & ``none'',{\bf ``smeft''}, ``user'' & Decides which
parametrization is used to describe interaction vertices - with $Z_X$
normalisation constants in an unexpanded form (``none''), using
``default'' SMEFT parameters (``smeft'') or user-defined set of
parameters (``user'') (see Sec.~\ref{sec:outpar} and examples in
Figs.~\ref{fig:vert_noexp}, \ref{fig:vert_smeft}, \ref{fig:vert_gf}).
\\[3mm]
InteractionFile & {\it filename} & Name of the file with mass basis
Lagrangian and vertices generated by {\tt SMEFTOutput} routine.
Default: {\tt output/smeft\_feynman\_rules.m}\\[3mm]
ModelFile & {\it filename} & Name of the model file containing SMEFT
parameters in mass basis generated by {\tt SMEFTOutput} routine.
Default: {\tt output/smeft\_par\_MB.fr}\\[3mm]
Include4Fermion & False, {\bf True} & 4-fermion vertices are not fully
supported by \frules - for extra safety calculations of them can be
switched off by setting this option to False.\\[3mm]
IncludeBL4Fermion & {\bf False}, True & Baryon and lepton number
violating 4-fermion vertices can be in principle evaluated by {\tt
  FeynRules}, but including them may lead to compatibility problems
with other codes - e.g.  MadGraph 5 reports errors if such vertices
are present in UFO file.  Thus in \sfr evaluation of such vertices is
by default switched off.  Set this option to True to include them.
\end{tabular}
\end{mdframed}
\end{center}
\caption{Options of {\tt SMEFTInitializeMB} routine, with default
  values marked in boldface.\label{tab:mbinit}}
\end{table}

\subsubsection{WCxf input and output}
\label{sec:wcxf}

Translation between \frules model files and WCxf format is done by the
functions {\tt SMEFTToWCXF} and {\tt WCXFToSMEFT}.  They can be used
standalone and do not require loading \frules and calling first {\tt
  SMEFTInitializeMB} routine to work properly.

Exporting numerical values of Wilson coefficients of operators in the
WCxf format is done by the function:

\medskip

{\tt SMEFTToWCXF[ SMEFT\_Parameter\_File, WCXF\_File, {\it FirstLetter
      $\to$ SMEFT\$MB ]} }

\medskip

\noindent where the arguments {\tt SMEFT\_Parameter\_File, WCXF\_File}
define the input model parameter file in the \frules format and the
output file in the WCxf JSON format, respectively.  Option {\tt
  FirstLetter} denote the first letter of names of WCs in a parameter
file and needs to be initialised only if it differs from variable {\tt
  MBFirstLetter} in Table~\ref{tab:init}.  The created JSON file can
be used to transfer numerical values of Wilson coefficients to other
codes supporting WCxf format.
Note that in general, the \frules model files may contain different
classes of parameters, according to the {\tt Value} property defined
to be a number (real or complex), a formula or even not defined at
all.  Only the Wilson coefficients with {\tt Value} defined to be a
number are transferred to the output file in WCxf format.

Conversely, files in WCxf format can be translated to \frules
parameter files using two routines:

\bigskip

\noindent {\tt ReadWCXFInput[ WCXF\_File, {\it Options} ]}\\
{\tt WCXFToSMEFT[ SMEFT\_Parameter\_File, {\it Options}] }

\bigskip

\noindent where {\tt ReadWCXFInput} reads values of WC from an input
file in the WCxf format and {\tt WCXFToSMEFT} creates parameter model
file for \frules which contain all necessary entries, including, apart
from WCs, also the definitions and numerical values of ``default'' and
``user-defined'' SMEFT input parameters.  The allowed options for both
routines defined in Table~\ref{tab:wcxf2}.

\begin{table}[htb!]
\begin{mdframed}[backgroundcolor=lightgray]  
\begin{tabular}{lp{27mm}p{90mm}}
Option & Allowed values & Description \\[2mm]
\hline\\
Operators & default: all operators & List with subset of Wilson
coefficients to be included in the SMEFT parameter file ({\tt
  ReadWCXFInput} only) \\[3mm]
RealParameters & False, True & Decides if only real values of Wilson
coefficients given in WCxf file are included in SMEFT parameter file.
The default value of this option is the same as set in the routine
SMEFTInitializeModel, see Table~\ref{tab:init}.  \\[3mm]
OverwriteTarget & {\bf False}, True & If set to True, target file is
overwritten without warning \\[3mm]
\end{tabular}
\end{mdframed}  
\caption{Options of {\tt ReadWCXFInput} and {\tt WCXFToSMEFT}
  routines.  Default values are marked in boldface.  Options
  RealParameters and OverwriteTarget affect only {\tt WCXFToSMEFT}.
\label{tab:wcxf2}}
\end{table}

\subsubsection{Latex output}
\label{sec:latex}

\sfr provides a dedicated Latex generator (not using the generic
\frules Latex export routine).  Its output has the following
structure:
\begin{itemize}
\item For each interaction vertex, the diagram is drawn, using the
  {\tt axodraw} style~\cite{Vermaseren:1994je}.  Expressions for
  Feynman rules are displayed next to corresponding diagrams.
\item In analytical expressions, all terms multiplying a given Wilson
  coefficient are collected together and simplified.
\item Long analytical expressions are automatically broken into many
  lines using {\tt breakn} style (this does not always work perfectly
  but the printout is sufficiently readable).
\item Latex output can be generated only for vertices expressed in
  terms of ``default'' SMEFT parameters, with $Z_X$ constants expanded
  in terms of WCs or kept as symbols (corresponding to options
  ``smeft'' or ``none'' in Tables~\ref{tab:mbinit}
  and~\ref{tab:latex}).  This is because the simplification of Latex
  formulae is optimised for such particular parametrizations, vertices
  calculated in terms of completely general ``user-defined'' parameter
  set may not be well readable.
\item Only terms up to maximal dimension 6 are included in Latex
  output.  Again, as above, this is because including higher order
  terms leads in most cases to lengthy and not very readable
  expressions.

\end{itemize}
Latex output is generated by the function:

\bigskip

{\tt SMEFTToLatex[ {\it Options} ] }

\bigskip

\noindent with the allowed options listed in Table~\ref{tab:latex}.
The function {\tt SMEFTToLatex} assumes that the variables listed in
Table~\ref{tab:variables} are initialised,thus it should be called
after reloading the mass basis Lagrangian with the {\tt
  SMEFTInitializeMB} routine, see Sec.~\ref{sec:interface}.

\begin{table}[htb!]
\begin{center}
\begin{mdframed}[backgroundcolor=lightgray]  
\noindent \begin{tabular}{lp{27mm}p{95mm}}
Option name & Allowed values & Description \\[2mm]
\hline\\
Expansion & {\bf ``none''}, ``smeft'' & Decides which parametrization
is used to describe interaction vertices - with $Z_X$ normalisation
constants in an unexpanded form (``none'') or using default SMEFT
parameters (``smeft'') (see discussion in Sec.~\ref{sec:outpar} and
examples in Figs.~\ref{fig:vert_noexp},
\ref{fig:vert_smeft},\ref{fig:vert_gf}).\\[3mm]
FullDocument & False, {\bf True} & By default a complete document is
generated, with all headers necessary for compilation.  If set to
False, headers are stripped off and the output file can be, without
modifications, included into other Latex documents.  \\[3mm]
ScreenOutput & {\bf False}, True & For debugging purposes, if set to
True the Latex output is printed also to the screen.
\end{tabular}
\end{mdframed}
\end{center}
\caption{Options of {\tt SMEFTToLatex} routine, with default values
  marked in boldface.\label{tab:latex}}
\end{table}

Latex output is stored in {\tt output/latex} sub-directory, split into
smaller files, each containing one primary vertex.  The main file is
named {\tt smeft\_feynman\_rules.tex}.  The style files necessary to
compile Latex output are supplied with the \sfr distribution.

Note that the correct compilation of documents using ``axodraw.sty''
style requires creating an intermediate Postscript file.  Programs
like {\it pdflatex} producing directly PDF output will not work
properly.  One should instead run in terminal in the correct directory
e.g.:

\medskip

{\tt

  latex smeft\_feynman\_rules.tex

  dvips smeft\_feynman\_rules.dvi

  ps2pdf smeft\_feynman\_rules.ps

}

\medskip

\noindent or equivalent set of commands, depending on the Latex
package used.

The {\tt smeft\_feynman\_rules.tex} does not contain analytical
expressions for five and six gluon vertices.  Such formulae are very
long (multiple pages, hard to even compile properly) and not useful
for hand-made calculations.  If such vertices are needed, they should
be rather directly exported in some other formats, as described in the
next subsection.

Other details not printed in the Latex output, such as, the form of
field propagators, conventions for parameters and momenta flow in
vertices (always incoming), manipulation of four-fermion vertices with
Majorana fermions \textit{etc}, are explained thoroughly in the
Appendices A1--A3 of ref.~\cite{Dedes:2017zog}.

\subsubsection{FeynArts and analytical calculation tests}
\label{sec:feynarts}

After calling the initialisation routine, {\tt SMEFTInitializeMB}, one
can generate output formats supported by native \frules interfaces, in
particular one can export SMEFT interactions and parameters to files
which could be imported by FeynArts (another especially important
format, UFO, is discussed separately in the next section).  For the
descriptions of the available output formats and commands used to
produce them, users should consult the \frules manual.  For instance,
to generate FeynArts output for the full mass basis Lagrangian, one
could call:
  
\medskip

\noindent {\tt WriteFeynArtsOutput[ SMEFT\$MBLagrangian, {\it Output
      $\to$ "output/FeynArts", \ldots}] }

\medskip

It is important to note that \frules interfaces like FeynArts (or UFO
described in Sec.~\ref{sec:ufo}), generate their output starting from
the level of SMEFT mass basis Lagrangian.  Thus, options of {\tt
  SMEFTInitializeModel} function like {\tt MajoranaNeutrino} and {\tt
  Correct4Fermion} (see Table~\ref{tab:init}) have no effect on output
generated by the interface routines.  As explained in
Sec.~\ref{sec:init} they affect only the expressions for Feynman rules
in {\tt FeynRules}/Mathematica format (which are also used to generate
Latex output file).

One should also note that \frules interfaces sometimes seem to be
``non-commuting''.  For example, calling FeynArts export routine
first, may lead to errors in subsequent execution of UFO interfaces
(like signalling problems with incorrect handling of vertices
containing explicit $\sigma^{\mu\nu}$ Dirac matrices or issues with
colour indices of SU(3) group structure constants), while the routines
called in opposite order are both working properly.  Therefore, it is
safer to generate one type of {\tt FeynRules}-supported output format
at a time and reinitialise model in mass basis if more output types
should be produced (WCxf and Latex generators does not suffer from
such issues and can be safely used together with others).

Finally, we have tested that our Feynman rules communicate properly
with {\tt FeynArts}.  An example of a non-trivial physics test we
performed is the following: we used the programs' chain {\tt SmeftFR
  $\to$ FeynArts $\to$ FormCalc} and calculated matrix elements for
longitudinal vector boson scattering processes, $V_L V_L\to V_L V_L$
with $V=W^\pm, Z$ at tree level with the full set of $d=6$ operators.
According to the Goldstone-Boson-Equivalence Theorem
(GBET)~\cite{Cornwall:1974km, Vayonakis:1976vz, Lee:1977eg,
  Chanowitz:1985hj}, at high energy this should be equal to the matrix
elements for the Goldstone Boson scattering processes $G G \to G G$
where, $G=G^\pm, G^0$ which should only contain WCs associated to
operators with powers of pure Higgs field $\varphi$ and its
derivatives.  All other, and there are many, WCs cancel out
non-trivially in all input {\tt ``user''} schemes employed by \sfr v3.
Similarly, we have also checked the validity of GBET (at tree level)
for $V_L V_L\to V_L V_L$ by including $d=6$ and $d=8$ operators
involving the full set of pure Higgs boson operators and its
derivatives.

It is perhaps instructive to provide one more test example for
  the dimension-8 operators: the positivity inequality constraints on
  WCs, see e.g.~\cite{Remmen:2019cyz,Yamashita:2020gtt}.  According to
  analyticity of the amplitude, the Froissart bound, and the optical
  theorem, for any elastic 2-to-2 scattering amplitude $\mathcal{M}(i
  j \to i j)$ of SM particles $i$ and $j$, the second derivative w.r.t
  the forward amplitude is positive semi-definite, i.e.,
%%%%%%%%%%%
\begin{equation}
    \frac{d^2}{ds^2} \, \mathcal{M} (i j \to i j) (s,t=0) \, \ge 0 \;,
    \label{eq:posi}
\end{equation}
%%%%%%%%%%%
where $s,t$ are the Mandelstam variables. 

By power counting, dimension-8 operators $Q_{\varphi^4
  D^4}^{(1,2,3)}$, potentially affect the matrix elements between the
Higgs ($h$) and the longitudinal components of the vector bosons
($Z_L$ and/or $W^\pm_L$), by a factor $s^2/\Lambda^4$. This can be
verified easily by using the {\tt FeynArt}-output of \sfr and
calculate the amplitudes with {\tt FormCalc}.  Then, application of
\eqref{eq:posi} to the relevant matrix elements of the processes
below, results in
%%%%%%%%%%%%%
\begin{eqnarray}
h h \to h h &\quad & \Longrightarrow \quad C_{\varphi^4 D^4}^{(1)} +
C_{\varphi^4 D^4}^{(2)} + C_{\varphi^4 D^4}^{(3)} \ge
0\;, \label{eq:p1} \\
Z_L h \to Z_L h &\quad & \Longrightarrow \quad C_{\varphi^4 D^4}^{(2)}
\ge 0\;, \label{eq:p2}\\
W_L^+ h \to W_L^+ h &\quad & \Longrightarrow \quad C_{\varphi^4
  D^4}^{(1)} + C_{\varphi^4 D^4}^{(2)} \ge 0 \;. \label{eq:p3}
\end{eqnarray}
%%%%%%%%%
All other longitudinal vector boson elastic scattering amplitudes
satisfy the above inequalities.  For example, applying \eqref{eq:posi}
to the amplitude $\mathcal{M}(W_L^+ W_L^+ \to W_L^+ W_L^+)$ gives
$C_{\varphi^4 D^4}^{(1)} + 2 \, C_{\varphi^4 D^4}^{(2)} + C_{\varphi^4
  D^4}^{(3)} \ge 0$, which is trivially satisfied by the inequalities
\eqref{eq:p1}-\eqref{eq:p3}. The above results are in agreement with
ref.~\cite{Remmen:2019cyz} and checked to be independent of the
input-parameter-schemes used by \sfr.  

Finally, several checks using Feynman Rules from \sfr with {\tt
  FormCalc} or {\tt FeynCalc} or by hand of various Ward-Identities
have been performed, and we always found agreement.

\subsubsection{UFO format and MadGraph 5 issues}
\label{sec:ufo}

Correct generation of UFO format requires more care.  UFO format
requires an extra parameter, ``interaction order'' (IO), to be
assigned to all couplings, to help Monte-Carlo generators like
MadGraph 5 decide the maximal order of diagrams included in amplitude
calculations.  It is customary to assign QED IO$=-1$ to Higgs boson
VEV, $v$, as it is numerically a large number and multiplying by $v$
can effectively cancel the suppression from smaller Yukawa or gauge
couplings.  In the SM, where all couplings are maximum dimension-4,
such procedure never leads to total negative IO for any vertex.
Unfortunately, in SMEFT some vertices are proportional to higher $v$
powers and technically can have negative total ``QED'' interaction
order, generating warnings when the UFO file is imported to MadGraph
5.  However, all such vertices have simultaneously another type of IO
assigned, ``NP=0,1,2'', defining their EFT order (which is
$1/\Lambda^{2\, \mathrm{NP}}$).  The ``NP'' order is sufficient for MC
generators to truncate the amplitude in a correct way, thus negative
``QED'' IO warnings can be ignored for such vertices.  To avoid
complications, \sfr v3 by default performs post-processing on UFO
output files, removing ``QED'' IOs from all vertices proportional to
WCs of higher dimension operators.  Such post-processing can be
switched off by setting the relevant option as described below.

Instead of {\tt FeynRules}'s {\tt WriteUFO} command, in \sfr v3 the
UFO output format can be generated by calling the routine:

\bigskip

\noindent {\tt SMEFTToUFO[ Lagrangian, {\it Options} ]}

\bigskip

\noindent with options defined in Table~\ref{tab:ufo}.  By default,
argument {\tt Lagrangian} should be set to variable named {\tt
  SMEFT\$MBLagrangian}, unless the user prefers to generate only
interaction for some sub-sector of the theory, then it can be one of
the variables defined in Table~\ref{tab:variables}, with obvious name
replacements like {\tt LeptonGaugeVertices $\to$
  LeptonGaugeLagrangian} etc.

One should note that some Monte-Carlo generators like MadGraph 5
support only real parameters, thus to generate UFO output working
properly one should use option {\tt RealParameters $\to$ True} when
calling {\tt SMEFTInitializeMB} routine.  Also, MadGraph 5 has some
hard coded names for QED and QCD coupling constants ({\tt ee, aEWM1,
  aS}).  For compatibility, \sfr v3 preserves those names,
independently of how the ``user-defined'' input parameters are named
(e.g., whatever is the name of the variable defining the strong
coupling constant, it is always copied to {\tt aS} used by MadGraph 5,
and similarly for other ``special'' variables).  If necessary for
compatibility with other codes, more such ``special'' variable names
could be added to the \sfr, editing the routine {\tt
  UpdateSpecialParameters} in the file {\tt smeft\_parameters.m}.

\begin{table}[htb!]
\begin{center}
\begin{mdframed}[backgroundcolor=lightgray]  
\noindent \begin{tabular}{lp{30mm}p{95mm}}
Option name & Allowed values & Description \\[2mm]
\hline\\
Output & {\bf ``output/UFO''} & default UFO output sub-directory, can
be modified to other user-defined location.\\[3mm]
CorrectIO & False, {\bf True} & By default only ``NP'' interaction
order parameter is left in vertices containing WCs of higher order
operators.  By setting this option to ``False", preserves all IOs
generated by native \frules UFO interface \\[3mm]
AddDecays & {\bf False}, True & UFO format can contain expressions for
2-body decays, switched off by default.
\end{tabular}
\end{mdframed}
\end{center}
\caption{Options of {\tt SMEFTToUFO} routine, with default values
  marked in boldface.\label{tab:ufo}}
\end{table}

If four-fermion vertices are included in SMEFT Lagrangian, the UFO
generator produces warning messages of the form (similar warnings may
appear also when using other \frules output routines):

\bigskip
\red{{\it Warning: Multi-Fermion operators are not yet fully
    supported!}}
\bigskip

Therefore, although in our experience it seems to work properly, the
output for four-fermion interactions in UFO or other formats must be
treated with care and limited trust --- performing appropriate checks
is left to users' responsibility.

Implementation in \frules of baryon and lepton number (BL) violating
four-fermion interactions, with charge conjugation matrix appearing
explicitly in vertices, is even more problematic.  Thus, for safety in
the current \sfr v3 such terms are by default not included in {\tt
  SMEFT\$MBLagrangian} variable, unless the option {\tt
  IncludeBL4Fermion} in {\tt SMEFTInitializeMB} routine is explicitly
set to {\tt True}.  In such case, FeynArts output seems to work for
such BL-violating vertices, but MadGraph 5 displays warnings that they
are not yet supported and aborts process generation.

We have tested that \sfr works properly with MadGraph5.  In
particular, we ran without errors test simulations in MadGraph5 v3.4.1
using UFO model files produced by \sfr v3. Furthermore, we performed
several types of numerical cross-checks against already existing
codes:
\begin{itemize}
\item we compared cross-sections for various processes obtained with
  \sfr against the results obtained with {\tt SMEFT@NLO} package up to
  terms of $\mathcal{O}(\Lambda^{-2})$ (note that {\tt SMEFT@NLO},
  {\tt Dim6Top} and {\tt SMEFTsim} have been formally validated up to
  this order~\cite{Durieux:2019lnv}, so it is sufficient to compare
  with only one of these codes),
\item we compared matrix elements for various processes obtained with
  \sfr against the results obtained with {\tt SMEFTsim} package up to
  terms of $\mathcal{O}(\Lambda^{-2})$, testing all implemented
  dimension 6 operators (apart from $B$- and $L$- violating ones),
\item we compared matrix elements for various processes obtained with
  \sfr against the results obtained with the code based on
  \cite{_boli_2016} (available at
  \url{https://feynrules.irmp.ucl.ac.be/wiki/AnomalousGaugeCoupling})
  up to terms of $\mathcal{O}(\Lambda^{-4})$, testing all operators
  considered in \cite{_boli_2016},
\end{itemize}
finding a very good agreement in each case.

%%%%%%%%%%%%
\begin{table}[htb!]
\centering {\small
\begin{tabular}{|c|c|c|c|}
\hline
& {\tt SMEFT@NLO} $\mathcal{O}(\Lambda^{-2})$& {\tt SmeftFR}
$\mathcal{O}(\Lambda^{-2})$ & {\tt SmeftFR}
$\mathcal{O}(\Lambda^{-4})$ \\
\hline
\multicolumn{4}{|c|}{$\mu^+ \mu^- \rightarrow t \bar{t}$} \\[1mm]
\hline
SM & $0.16606\pm 0.00026$ & $0.16608\pm 0.00024$ & - \\
\hline
$C_{uW}^{33}$ & $0.41862 \pm 0.00048$ & $0.41816 \pm 0.00047$ & - \\
\hline
$C_{\varphi u}^{33}$ & $0.16725 \pm 0.00027$ & $0.16730 \pm 0.00025$ &
- \\
\hline
$C_{l u}^{2233}$ & $6.488 \pm 0.016$ & $6.491 \pm 0.014$ & - \\
\hline
$C_{\varphi WB}$ & $ 0.21923 \pm 0.00032 $ & $0.21940 \pm 0.00030$ &
$0.22419 \pm 0.00030 $ \\
\hline
$C_{\varphi D}$ & $0.18759 \pm 0.00030$ & $0.18759 \pm 0.00027 $ & $
0.18829 \pm 0.00027$ \\
\hline
\multicolumn{4}{|c|}{$\gamma \gamma \rightarrow t \bar{t}$} \\[1mm]
\hline
SM & $0.0037498\pm 0.0000050$ & $0.0037498\pm 0.0000050$ & - \\
\hline
$C_{uW}^{33}$ & $0.008229 \pm 0.000012$ & $0.008235 \pm 0.000012$ & -
\\
\hline
$C_{\varphi WB}$ & $ 0.0053056\pm 0.0000086$ & $0.0053056\pm
0.0000086$ & $0.0055809 \pm 0.0000090 $ \\
\hline
$C_{\varphi D}$ & $0.0045856 \pm 0.0000061$ & $0.0045895 \pm
0.0000064$ & $0.0045882 \pm 0.0000069$ \\
\hline
\multicolumn{4}{|c|}{$c \bar{c} \rightarrow t \bar{t}$} \\[1mm]
\hline
SM & $0.9553 \pm 0.0017$ & $0.9511 \pm 0.0023$ & - \\
\hline
$C_{uG}^{33}$ & $1.1867 \pm 0.0023$ & $1.1854 \pm 0.0021$ & - \\
\hline
$C_{uW}^{33}$ & $0.9641\pm0.0018$ & $0.9599 \pm 0.0024$ & - \\
\hline
$C_{\varphi u}^{33}$ & $0.9555 \pm 0.0017$ & $0.9513\pm 0.0023$ & - \\
\hline
$C_{\varphi q3}^{33}$ & $0.9558 \pm 0.0017$ & $0.9515 \pm 0.0023$ & -
\\
\hline
$C_{q u 1}^{2233}$ & $1.0111 \pm 0.0018$ & $1.0059 \pm 0.0015$ & - \\
\hline
$C_{\varphi WB}$ & $0.9568 \pm 0.0018$ & $0.9520\pm0.0018$ & $0.9522
\pm 0.0018$ \\
\hline
$C_{\varphi D}$ & $0.9558 \pm 0.0017$ & $0.9511 \pm 0.0018$ & $0.9511
\pm 0.0018$ \\
\hline
\multicolumn{4}{|c|}{$pp\rightarrow  t \bar{t}$ } \\[1mm]
\hline
SM & $510.35 \pm 0.72$ & $510.46 \pm 0.68$ & - \\
\hline
$C_{uG}^{33}$ & $664.33 \pm 1.16$ & $666.34 \pm 0.90$ & $671.08 \pm
0.97$ \\
\hline
$C_{uW}^{33}$ & $ 510.63 \pm 0.70$ & $510.70\pm 0.80$ & - \\
\hline
$C_{\varphi u}^{33}$ & $ 510.37 \pm 0.72$ & $ 510.47 \pm 0.68$ & - \\
\hline
$C_{\varphi q3}^{33}$ & $510.39 \pm 0.72$ & $510.65 \pm 0.80$ & - \\
\hline
$\sum_{i=1,2} C_{q u 1}^{ii33}$ & $516.31 \pm 0.58$ & $516.14 \pm
0.64$ & - \\
\hline
%
%$C_{q u 1}^{2233}$ & $514.31 \pm 0.70 $ & $510.59 \pm 0.67$ & $509.03
%\pm 0.79$ \\
%\hline
%
$C_{\varphi WB}$ & $510.49 \pm 0.68$ & $510.52 \pm 0.71$ & $508.94 \pm
0.79$ \\
\hline
$C_{\varphi D}$ & $510.38 \pm 0.72$ & $510.47\pm 0.68$ & $508.89 \pm
0.79$ \\
\hline
\end{tabular}
}%end of \small
\caption{Cross-sections (in pb) obtained using {\tt
    MadGraph5} with UFO models provided by {\tt SMEFTatNLO} at the
  $\mathcal{O}(\Lambda^{-2})$ order of the EFT expansion and \sfr at
  the $\mathcal{O}(\Lambda^{-2})$ and $\mathcal{O}(\Lambda^{-4})$
  orders of the EFT expansion for a chosen set of processes and SMEFT
  operators.  An empty cell indicates that no $\mathcal{O}(\Lambda^{-4})$
  terms appear in the amplitude.}
\label{table:comparisonUFO}
\end{table}
%%%%%%%%%%%%%%%%%
For all comparisons which we performed we have used the
$(G_F,M_W,M_Z,M_H)$ input parameter scheme (option {\tt InputScheme
  $\to$ "GF"} in {\tt SMEFTInitializeModel} routine) with values of
input parameters set to central values given in
ref.~\cite{ParticleDataGroup:2020ssz} {(unless explicitly stated
  otherwise below)}. In addition, CKM and PMNS matrices were
approximated by unit matrices.

For cross-sections comparison, all particle widths, fermion masses and
Yukawa couplings, except for the top quark, were assumed to be zero.
Each cross section was calculated assuming that all but one Wilson
coefficients were set to zero and the non-vanishing one (displayed in
the left column of Table~\ref{table:comparisonUFO}) had the value of
$\left |\frac{C_i}{\Lambda^2}\right|=10^{-6}$ GeV$^{-2}$, while its
sign was always chosen to increase $\mathcal{O}(\Lambda^{-2})$ cross
section w.r.t. SM.  The results are summarised in the 2nd and 3rd
column of Table~\ref{table:comparisonUFO}.  As one can see,
differences between both codes at the $\mathcal{O}(\Lambda^{-2})$
level never exceed $1\%$.

The novel capability of \sfr v3 is the consistent inclusion of
$O(1/\Lambda^4)$ terms in the interaction vertices.  Therefore, \sfr
v3 is able to {\it exactly} calculate dimension-6 squared terms in the
amplitude.  For completeness, we have checked the impact of such
$\mathcal{O}(\Lambda^{-4})$ terms for the same processes.  The
corresponding cross sections can be found in the 4th column of the
Table ~\ref{table:comparisonUFO}. The effect of higher order
contributions is visible albeit small for the chosen small input
values of WCs. However, in another example using \sfr with a large
coefficient $C_W$ displayed in Table~7 of
ref.~\cite{Aebischer:2023irs}, the effect of dimension-6-squared terms
on cross-section for $W$-boson scattering can be different by factors
of thousand!

We have used similar procedure for matrix elements comparison. Once
again each matrix element was calculated assuming that all but one
Wilson coefficients were set to zero and the non-vanishing one had the
value of $\frac{C_i^6}{\Lambda^2}=10^{-6}$ GeV$^{-2}$ for dimension-6
and $\frac{C_i^8}{\Lambda^2}=10^{-11}$ GeV$^{-4}$ for dimension-8
coefficients.  We obtained almost identical results from {\tt
  SMEFTsim} or \url{AnomalousGaugeCoupling} and \sfr for all of the
studied processes, with the differences not exceeding $0.1$\%, usually
being much smaller.  As the number of compared processes is large, we
do not include here the detailed comparison tables, they can be found
on the \sfr homepage \url{https://www.fuw.edu.pl/smeft}.

\subsection{Potential problems and optional further \sfr extensions}
\label{sec:issues}

As already mentioned before, SMEFT itself is a very complicated model
even at the level of Lagrangian construction. A transition amplitude
calculations within SMEFT can easily increase the complexity of
required analytical and numerical computations beyond the capability
of humans or computers.  Therefore, to remain within reasonable limits
of time and effort required for a given analysis, it is strongly
advised to generate necessary SMEFT interactions only for a subset of
operators relevant to a chosen problem, a task for which \sfr was
specifically designed for.

We performed number of tests to estimate the CPU time required to run
the code for various initial \sfr v3 setups. Deriving Feynman rules in
Mathematica/FeynRules format up to dimension-6 terms and for complete
dimension-6 SMEFT Lagrangian (i.e. including 60 independent operators
in Warsaw basis with fully general flavour structure and all numerical
values of parameters initialised) takes about an hour on typical PC
computer (depending on its speed of course), more if interaction
vertices need to be expressed in terms of user-defined input
parameters.  Exporting Feynman rules to UFO or other formats is more
time consuming, can take few or more hours.  Including also all
dimension-6 squared terms and the full set of bosonic dimension-8
operators at once does not seem to be feasible at all, as the
computations can exhaust even large computer memory and/or human
patience. For such calculations, choosing the subset of SMEFT operators
is unavoidable.

Further problems related to complexity of SMEFT interactions,
especially at the full di\-men\-sion-8 level, may arise when importing
the \sfr output to other public codes. In particular, in some cases we
encountered difficulties when generating SMEFT processes with
MadGraph5 - the Feynman rules in UFO file generated by \sfr contained
such a lengthy expressions for interaction vertices that MadGraph
internal compiler was unable to process them in a correct way and
reported errors.  Again, such issues could be solved (apart from using
different Fortran or C compiler!) by limiting the number of included
operators and/or decreasing the required order of EFT expansion to
dimension-6 only.

\section{Sample programs}
\label{sec:sample}

After setting the variable {\tt \$FeynRulesPath} to the correct value,
in order to evaluate mass basis SMEFT Lagrangian and analytical form
of Feynman rules for some sample set of dimension-6 and 8 operators
one can use the following sequence of commands:

\bigskip

{\tt

\noindent SMEFT\$MajorVersion      = "3";\\
SMEFT\$MinorVersion      = "01";\\
SMEFT\$Path              = FileNameJoin[\{\$FeynRulesPath, "Models", "SMEFT\_" <> \\
\hspace*{4cm} SMEFT\$MajorVersion <> "\_" <> SMEFT\$MinorVersion\}];\\
\\
Get[ FileNameJoin[\{\$FeynRulesPath,"FeynRules.m"\}] ];\\
Get[ FileNameJoin[\{ SMEFT\$Path, "code", "smeft\_package.m"\}] ];\\
\\
OpList6 = \{}{\tt "phi"}, {\tt "phiBox"}, {\tt "phiD"}, {\tt
  "phiW"}, {\tt "phiWB"}, {\tt "eB"}, {\tt "uW"}, {\tt "dphi"}, {\tt
  "ll"}{\tt\};\\
OpList8 = \{}{\tt "phi8"}, {\tt "phi4n1"}, {\tt "phi4n3"}{\tt\};\\
OpList = Join[OpList6, OpList8]; \\[2mm]
\begin{tabular}{ll}
SMEFTInitializeModel[ &  Operators -> OpList,\\
& Gauge -> Rxi,\\
& WCXFInitFile -> "wcxf\_input\_file\_with\_path.json"\\
&  ExpansionOrder -> 1,\\
&  InputScheme -> "GF",\\
&  CKMInput -> "yes",\\
& RealParameters -> True,\\
&  MaxParticles -> 4,\\  
&  MajoranaNeutrino -> True,\\  
& Correct4Fermion -> False ]; 
\end{tabular}
\\[2mm]
SMEFTLoadModel[ ];\\
SMEFTFindMassBasis[ ];\\
SMEFTFeynmanRules[ ];\\
SMEFTOutput[ ];
}

\bigskip

\noindent or alternatively rerun the supplied programs: the notebook
          {\tt SmeftFR-init.nb} or the text script {\tt
            smeft\_fr\_init.m}.

After running the sequence of commands listed above, interaction
vertices in different parametrizations become available and can be
displayed on screen or used in further calculations.  For example, the
Higgs-photon-photon vertex for the fields in mass basis can be
extracted in different schemes by using the commands:

\bigskip

{\tt

\noindent Print["Higgs-photon-photon vertex in "none" scheme:
  ",\\ SelectVertices[GaugeHiggsVertices, SelectParticles -> {H, A,
      A}]];\\[2mm]
SMEFTExpandVertices[Input -> "smeft", ExpOrder -> 2];\\
Print["Higgs-photon-photon vertex in "smeft" scheme:
  ",\\ SelectVertices[GaugeHiggsVerticesExp, SelectParticles -> {H, A,
      A}]];\\[2mm]
SMEFTExpandVertices[Input -> "user", ExpOrder ->
  2];\\ Print["Higgs-photon-photon vertex in "user" scheme:
  ",\\ SelectVertices[GaugeHiggsVerticesExp, SelectParticles -> {H, A,
      A}]]; }

\bigskip

As described before, Latex, WCxf, UFO and FeynArts formats can be
exported after rerunning first {\tt SmeftFR-init.nb} or equivalent set
of commands generating file {\tt smeft\_feynman\_rules.m} containing
the expressions for the mass basis Lagrangian.  Then, the user needs
to start a new \textit{Mathematica} kernel and rerun the notebook file
{\tt SmeftFR-interfaces.nb} or the script {\tt
  smeft\_fr\_interfaces.m}.  Alternatively, one can manually type the
commands, if necessary changing some of their options as described in
previous Sections:

\bigskip

{\tt
\noindent Get[ FileNameJoin[\{\$FeynRulesPath,"FeynRules.m"\}] ];\\
Get[ FileNameJoin[\{SMEFT\$Path, "code", "smeft\_package.m"\}] ];\\[2mm]
SMEFTInitializeMB[ Expansion->"user", Include4Fermion->True, ];\\[1mm]
SMEFTToWCXF[ SMEFT\$Path<>"output/smeft\_par\_MB.fr", \\
\hspace*{26mm} SMEFT\$Path<>"output/smeft\_wcxf\_MB.json" ];\\[1mm]
SMEFToLatex[ Expansion -> "smeft" ];\\[1mm]
SMEFTToUFO[ SMEFT\$MBLagrangian, CorrectIO -> True, Output -> $\ldots$
];\\[1mm]
WriteFeynArtsOutput[ SMEFT\$MBLagrangian, Output -> $\ldots$ ];\\

}

A step-by-step example on how to use \sfr v3 in practice is given in
ref.~\cite{Aebischer:2023irs}.

%%%%%%%%%%%%%%%%%%%%%%%%%
\section{Future Implementations}
\label{sec:future}
%%%%%%%%%%%%%%%%%%%%%%%%

There are various important implementations that have been left out
from the current version, \sfr v3, with the most pressing being the
inclusion of fermionic dimension-8 operators. For instance, the latter
have been proven recently~\cite{Degrande:2023iob} to provide dominant
effects in vector-boson production.
Unfortunately, including all such operators in full generality is
  difficult - they are numerous and implementing them correctly
  requires, comparing to pure bosonic case, taking into account their
  tensor structures in the flavour space, transformation properties
  under flavour rotations (necessary in transition to mass eigenstates
  basis), symmetry properties under flavour index permutations, etc.
  Apart from technical problems, computations involving large number
  of fermionic dimension-8 operators can exceed reasonable CPU running
  time and computer memory limits.

Nevertheless, as we have already mentioned, selected dimension-8
fermionic operators can be added to \sfr v3.  However, it requires
intervention in many parts of the code. For test purposes,
we were able to successfully add a sample of dimension-8 fermionic
operators to \sfr v3, and have documented the complete list of
required code changes. At present, such prescription is rather
complicated and requires knowledge of the internal code structure more
detailed than can be expected from most users, so we decided not to
include it in the current version of the manual. The file with the
detailed instructions on how to do that is available on the web page of
\sfr, \webpage.  If it is not sufficient, users interested in adding
dimension-8 fermionic operators to \sfr can contact the authors for
further help. We plan to include a simplified procedure of adding
higher order fermionic operators in the next version of \sfr.

%%%%%%%%%%%%%%%%%%%%%%%%%
\section{Summary}
\label{sec:summary}
%%%%%%%%%%%%%%%%%%%%%%%%

In recent years, SMEFT has become the standard framework for a
concrete, robust, organised, and fairly model independent way of
capturing physics beyond the SM.  Huge efforts among the high energy
community physicists, both theoretical and experimental, have been
devoted to understand how to precisely map experimental observable and
fit them onto the Wilson coefficients of the SMEFT Lagrangian in
eq.~\ref{Leff}.  Even deriving the Feynman rules - a straightforward
and most of the time effortless procedure in renormalizable theories -
is not trivial in SMEFT: The abundance of operators and associated
parameters, especially when climbing up in EFT-dimensionality, makes
the computer aid necessary, if not indispensable.

In this paper, we present a new version of a code, the \sfr v3,
previous versions of which had been tested in many work studies.  \sfr
v3 is able to express the SMEFT interaction vertices in terms of
chosen, predefined or user-defined, set of observable input
parameters, avoiding the need for reparametrizations required in
calculations when expressed in terms of the SM gauge, Yukawa and Higgs
coupling constants.
One of \sfr v3 main advantages is that, it can calculate SMEFT
interactions {\it \`a la carte} for user-defined subset of
dimension-5, 6 and 8 operators, selected to be relevant to scattering
matrix elements for observable (or observables) under scrutiny.  It
generates dynamically the corresponding \frules model files with the
minimal required content, in effect producing more compact analytical
formulae and significantly speeding up the numerical computations.
The SMEFT Feynman rules can be calculated by \sfr v3 in unitary and
$R_\xi$-gauges, following the procedure described in
ref.~\cite{Dedes:2017zog}.  A number of additional \sfr v3's options
is described in details in this paper.

The output of the package can be printed in Latex or exported in
various formats supported by {\tt FeynRules}, such as UFO, FeynArts,
{\it etc}.  Input parameters for Wilson coefficients used in \sfr v3
can communicate with WCxf format for further numerical handling.

We have also performed a number of analytical and numerical
consistency checks that came out from \sfr v3 calculations.
Analytically, for example, we checked that the produced Feynman rules
lead to correct non-trivial cancellations in Vector Boson Scattering
helicity amplitudes in our predefined input-parameter schemes, certain
Ward identities and positivity of combinations of dimension-8 Wilson
coefficients.  Numerically, we found very good agreement with other
codes, such as {\tt SMEFTsim} and {\tt SMEFT@NLO}, commonly used for
Monte-Carlo simulations in SMEFT.  Compared to those codes, \sfr v3
offers in addition several important improvements: the precision of
including consistently terms up to $O(1/\Lambda^4)$ (that is all
(dimension-6)$^2$ terms and the full set of terms linear in WCs of
bosonic dimension-8 operators), the physical input-parameter-schemes
not only for the gauge and Higgs sector but also for the flavour
sector by including SMEFT corrections to the CKM matrix, the inclusion
of the SMEFT neutrino sector, and inclusion of the Baryon and Lepton
number violating $d=6$ interaction vertices.

The current version of \sfr v3 code and its manual can be downloaded
from 
\begin{center}
\webpage
\end{center}

We believe that \sfr v3 is an important tool, facilitating the
computations within SMEFT from the theoretical Lagrangian level all
the way down to amplitude calculations required by the beyond the SM
physics experimental analyses.

\section*{Acknowledgements}

The work of MR was supported in part by Polish National Science Centre
under research grant
% JR grant
DEC-2019/35/B/ST2/02008.
The research of JR has received funding from the Norwegian Financial
Mechanism for years 2014-2021, under the grant no 2019/34/H/ST2/00707.
The research work of LT was supported by the Hellenic Foundation for
Research and Innovation (HFRI) under the HFRI PhD Fellowship grand
(fellowship number: 1588).  LT would like to thank the University of
Warsaw for hospitality and financial support during his stay there.
JR would like to thank the University of Ioannina and CERN for 
hospitality during his stays there.
AD would like to thank CERN for hospitality.
We would like to thank Dimitrios Beis for checking the SMEFT
contributions to CKM and the numerical output of \sfr with an
independent code.

%%%%%%%%%%%%%%%%%%%%%
\newpage
%%%%%%%%%%%%%%%%%%%%%%%

\appendix
\renewcommand{\thesection}{Appendix~\Alph{section}}
\renewcommand{\thesubsection}{\Alph{section}.\arabic{subsection}}
\renewcommand{\theequation}{\Alph{section}.\arabic{equation}}
\setcounter{table}{0}

\section{Input schemes for the electroweak sector}
\label{app:inp}

The electroweak sector parameters, $\bar{g}$, $\bar{g}^\prime$, $v$
and $\lambda$, after expansion in $1/\Lambda$-powers can be written in
the following form:
\bea
\bar{g} &=& \bar{g}_{SM}+\frac{1}{\Lambda^2}\bar{g}_{D6} +
\frac{1}{\Lambda^4}\bar{g}_{D8}\;,\nn
\bar{g}^\prime &=& \bar{g}^\prime_{SM}+\frac{1}{\Lambda^2}\bar{g}^\prime_{D6} +
\frac{1}{\Lambda^4}\bar{g}^\prime_{D8}\;,\nn
v &=& v_{SM}+\frac{1}{\Lambda^2}v_{D6}
+ \frac{1}{\Lambda^4}v_{D8}\;,\nn
\lambda &=& \lambda_{SM} + \frac{1}{\Lambda^2}\lambda_{D6} +
\frac{1}{\Lambda^4}\lambda_{D8}\;.
\label{eq:linearised}
\eea
where the exact form of ``SM'', ``D6'' and ``D8'' terms depends on the
chosen input scheme.  Below, we present relevant expressions for the
two most commonly used SMEFT input schemes, both included as
predefined routines in the \sfr v3 distribution.

\subsection{``GF'' input scheme}
\label{app:gf}

In this scheme Fermi constant $G_F$ (evaluated from the muon lifetime
measurement) and gauge and Higgs boson masses $M_Z, M_W, M_H$ are used
as the input parameters.  To relate them to quantities defined in
eq.~(\ref{eq:linearised}), let us first define the following
abbreviations
\bea
\Delta M&=& \sqrt{M_Z^2-M_W^2}\;,\nn
\mathcal{B}_6(C_{ll},C_{\varphi l3}) &=& -2(C_{ll}^{2112}-C_{\varphi l
  3}^{11} - C_{\varphi l 3}^{22})\;,\nn
\mathcal{B}_8(C_{ll},C_{\varphi l3},C_{\varphi l1})&=&
(C_{ll}^{2112})^2 + \frac{1}{4}(C_{le}^{2112})^2 -
2C_{ll}^{2112}C_{\varphi l 3}^{11} - 2 C_{ll}^{2112} C_{\varphi l
  3}^{22} \nn
&+& (C_{\varphi l 3}^{11})^2 + (C_{\varphi l 3}^{22})^2 + 4 C_{\varphi
  l 3}^{11}C_{\varphi l 3}^{22}\nn
&+&C_{\varphi l 1}^{21} C_{\varphi l 3}^{12} - C_{\varphi l 1}^{12}
C_{\varphi l 3}^{21} + C_{\varphi l 1}^{12} C_{\varphi l 1}^{21} -
C_{\varphi l 3}^{12} C_{\varphi l 3}^{21} \;.
\label{eq:abbrev}
\eea
Then one can express quantities in eq.~(\ref{eq:linearised}), as
{\small
\bea
v_{SM}&=&\frac{1}{2^{1/4}\sqrt{G_F}}\;,\nn
v_{D6}&=& \frac{v_{SM}}{4\sqrt{2}G_F}\mathcal{B}_6\;,\nn
v_{D8}&=&\frac{v_{SM}}{64G_F^2}(\mathcal{B}_6^2 +
8\mathcal{B}_8)\;,\\[3mm]
\bar{g}_{SM} &=& 2^{5/4}\sqrt{G_F}M_W\;,\nn
\bar{g}_{D6} &=& -\frac{\bar{g}_{SM} }{4\sqrt{2}
  G_F}\mathcal{B}_6\;,\nn
\bar{g}_{D8} &=& \frac{\bar{g}_{SM}}{64G_F^2} (\mathcal{B}_6^2 -
8\mathcal{B}_8)\;, \\[3mm]
\bar{g}^\prime_{SM}&=&2^{5/4}\sqrt{G_F}\Delta M^2\;,\nn
\bar{g}^\prime_{D6}&=&\frac{\bar{g}^\prime_{SM}}{4\sqrt{2}G_F \Delta
  M}
\left(- M_Z^2C_{\varphi D} - 4 M_W \Delta M C_{\varphi WB} - \Delta
M^2\mathcal{B}_6\right)\;,\nn
\bar{g}^\prime_{D8} &=& \frac{\bar{g}^\prime_{SM}}{16G_F^2\Delta
  M^2}\biggl[ - 2M_Z^2 (2 C_{\varphi^6 D^2} + \mathcal{B}_6 C_{\varphi
    D}) +\Delta M^2 (\mathcal{B}_6^2-8\mathcal{B}_8 - 16 C_{\varphi
    WB}^2)\biggr.\nn
&-&\biggl.  8 M_W \Bigl( 2M_W C_{W^2 \varphi^4}^{(3)} + 2\Delta M
  C_{WB\varphi^4}^{(1)} + \Delta M (\mathcal{B}_6+4C_{\varphi B}
  +4C_{\varphi W})C_{\varphi WB}\Bigr)\biggr] \;,\\[3mm]
\lambda_{SM} &=& \sqrt{2} G_F M_H^2 \;,\nn
\lambda_{D6} &= &\frac{\lambda_{SM}}{4G_F} \biggl[ \frac{6}{G_F
    M_H^2}C_\varphi - \sqrt{2}\Bigl(\mathcal{B}_6 + 4 C_{\varphi \Box}
  - C_{\varphi D}\Bigr)\biggr]\;,\nn
\lambda_{D8}&=&\frac{\lambda_{SM}}{16G_F^2}\biggl[\Bigl(\mathcal{B}_6^2
  - 4\mathcal{B}_8 - 8 C_{\varphi^6 \Box} + 2 C_{\varphi^2 D^2}\Bigr)
  + \frac{6\sqrt{2}}{G_F M_H^2}\Bigl(\mathcal{B}_6 C_\varphi + 2
  C_{\varphi 8}\Bigr)\biggr]\;.
\label{eq:GFscheme}
\eea
}

\subsection{``AEM'' input scheme}
\label{app:aem}

In this scheme input parameters for the electroweak sector are chosen
to be the electromagnetic coupling $\alpha_{em}$, and the gauge and
Higgs boson masses $M_Z, M_W, M_H$.  Using again the abbreviation
$\Delta M = \sqrt{M_Z^2-M_W^2}$, for the quantities defined in
eq.~(\ref{eq:linearised}), one has:

{\small
\bea
v_{SM} &=& \frac{M_W \Delta M }{M_Z\sqrt{\pi\alpha_{em}}}\;,\nn
v_{D6} &=& -\frac{\bar{g}_{SM} M_W^3}{4 \pi \alpha_{em} M_Z^2} \left(
M_W C_{\varphi D} + 4 \Delta M C_{\varphi WB}\right)\;, \nn
v_{D8} &=& \frac{v_{SM} M_W^5}{32 \pi^2 \alpha_{em}^2 M_Z^4}
\biggl[3M_W^3 C_{\varphi D}^2 - 4 M_W \Delta M^2 C_{\varphi^6 D^2} - 8
  (M_Z^2-5 M_W^2)\Delta M C_{\varphi D}C_{\varphi WB}\biggr.\nn
&+& 16\Delta M^2 \Bigl(4 M_W C_{\varphi WB}^2 -\Delta M
  C_{WB\varphi^4}^{(1)} +
  \frac{M_Z^2-2M_W^2}{M_W}C_{WB\varphi^4}^{(3)}\Bigr)\nn
&-& \biggl.32 \Delta M^3(C_{\varphi B}+C_{\varphi W}) C_{\varphi
    WB}\biggr]\;,\\[3mm]
\bar{g}_{SM} &=& \frac{2 M_Z \sqrt{\pi \alpha_{em}}}{\Delta M}\;, \nn
\bar{g}_{D6} &=& -v_{D6}\;,\nn
\bar{g}_{D8} &=& \frac{\bar{g}_{SM} M_W^5 }{32 \pi^2 \alpha_{em}^2
  M_Z^4} \biggl[-M_W^3C_{\varphi D}^2 + 4 M_W \Delta M^2 C_{\varphi^6
    D^2} + 8(M_Z^2-3M_W^2)\Delta M C_{\varphi D}C_{\varphi
    WB}\biggr.\nn
&-& 16\Delta M^2 \Bigl( 2M_W C_{\varphi WB}^2 -\Delta M
  C_{WB\varphi^4}^{(1)} +
  \frac{M_Z^2-2M_W^2}{M_W}C_{WB\varphi^4}^{(3)}\Bigr)\nn
&+& \biggl.  32\Delta M^3 (C_{\varphi B} + C_{\varphi W})C_{\varphi
    WB}\biggr]\;,\\[3mm]
\bar{g}^\prime_{SM} &=& \frac{2 M_Z\sqrt{\pi\alpha_{em}}}{M_W}\;,\nn
\bar{g}^\prime_{D6} &=& - \frac{\bar{g}^\prime_{SM} \Delta M^2
  M_W^2}{4 \pi \alpha_{em} M_Z^2}C_{\varphi D}\;,\nn
\bar{g}^\prime_{D8} &=& \frac{\bar{g}^\prime_{SM} M_W^4\Delta M^2}{32
  \pi^2 \alpha_{em}^2 M_Z^4} \biggl[(M_W^2+3 M_Z^2)C_{\varphi D}^2 -
  16\Delta M^2 C_{\varphi WB}^2 + 16M_W\Delta M C_{\varphi
    D}C_{\varphi WB}\biggr.\;\nn
&-& \biggl.  4\Delta M^2 \Bigl(C_{\varphi^6 D^2}+4 C_{W^2
    \varphi^4}^{(3)}\Bigr)\biggr]\;,\\[3mm]
\lambda_{SM} &=& \frac{\pi \alpha_{em}M_H^2 M_Z^2 }{\Delta M^2}\;, \nn
\lambda_{D6} &=& \frac{3\Delta M^2 M_W^2}{\pi \alpha_{em} M_Z^2}
C_\varphi - 2M_H^2 C_{\varphi \Box} + \frac{M_H^2 M_Z^2}{2\Delta M^2}
C_{\varphi D} +2 M_W\Delta M C_{\varphi WB} \;,\nn
\lambda_{D8} &=& \frac{M_W^2}{4\pi^2\alpha_{em}^2M_Z^4} \biggl[
  12M_W^2\Delta M^4 2C_{\varphi 8} - 6 M_W^3\Delta M^2 C_\varphi (M_W
  C_{\varphi D}+4 \Delta M C_{\varphi WB}) \biggr.\;\nn
&+& \pi\alpha_{em}M_Z^2 M_H^2 \Bigl(-4\Delta M^2 C_{\varphi^6\Box}
  +M_Z^2C_{\varphi^2D^2} +8M_W\Delta M(C_{\varphi B}+ C_{\varphi
    W})C_{\varphi WB}\Bigr.\;\nn
&+&\biggl.   \frac{2 M_W(M_Z^2-2M_W^2)}{\Delta M} C_{\varphi D}
  C_{\varphi WB} - 4 M_W^2 C_{\varphi WB}^2 \nn
&+&\Bigl.   4 M_W \Delta M C_{WB\varphi^4}^{(1)} - 4(M_Z^2-2M_W^2)
  C_{WB\varphi^4}^{(3)} \Bigr) \biggr] \;.
\label{appendix:AEM}
\eea

%end of \small
}

%%%%%%%%%%%%%%%%%%%%%%%%%%%%%%%%%%%%%%%%%%%%%%%%%%%%%%%%%%
\section{Operators and their naming used in \sfr }
\label{app:ops}
%%%%%%%%%%%%%%%%%%%%%%%%%%%%%%%%%%%%%%%%%%%%%%%%%%%%%%%%

All dimension-6 operators in Warsaw basis are given in
Table~\ref{tab:sfr3-dim6} (copied here for complementarity from
ref.~\cite{Grzadkowski:2010es}).  Naming of \sfr variables
corresponding to WCs of these operators is straightforward: each
variable name consists of subscripts identifying a given operator,
with obvious transcriptions of ``tilde'' symbol and Greek letters to
Latin alphabet. Operator names are represented by strings, to avoid
accidental use of similarly named variables for other purposes. For
example, one may include in {\tt OpList6} (list of dimension-6
operators, see examples in Sec.~\ref{sec:sample}):
\begin{align*}
    Q_{\varphi} &\rightarrow {\tt ``phi"}\\
    Q_{\varphi D} &\rightarrow {\tt ``phiD"}\\
    Q_{\varphi\Box} &\rightarrow {\tt ``phiBox"}\\
    Q_{\varphi\widetilde W} &\rightarrow {\tt ``phiWtilde"} \\
    Q_{lq}^{(3)} &\rightarrow {\tt ``lq3"}\\
    Q_{quqd}^{(8)} &\rightarrow {\tt ``quqd8"}\\
    ....  \nonumber
\end{align*}
The full list of all dimension-6 operators contains the following
entries:
{\small

\medskip
  
\noindent {\tt OpList6} = \{ {\tt "G"}, {\tt "Gtilde"}, {\tt "W"},
          {\tt "Wtilde"}, {\tt "phi"}, {\tt "phiBox"}, {\tt "phiD"},
          {\tt "phiW"}, {\tt "phiB"}, {\tt "phiWB"}, {\tt
            "phiWtilde"}, {\tt "phiBtilde"}, {\tt "phiWtildeB"}, {\tt
            "phiGtilde"}, {\tt "phiG"}, {\tt "ephi"}, {\tt "dphi"},
          {\tt "uphi"}, {\tt "eW"}, {\tt "eB"}, {\tt "uG"}, {\tt
            "uW"}, {\tt "uB"}, {\tt "dG"}, {\tt "dW"}, {\tt "dB"},
          {\tt "phil1"}, {\tt "phil3"}, {\tt "phie"}, {\tt "phiq1"},
          {\tt "phiq3"}, {\tt "phiu"}, {\tt "phid"}, {\tt "phiud"},
          {\tt "ll"}, {\tt "qq1"}, {\tt "qq3"}, {\tt "lq1"}, {\tt
            "lq3"}, {\tt "ee"}, {\tt "uu"}, {\tt "dd"}, {\tt "eu"},
          {\tt "ed"}, {\tt "ud1"}, {\tt "ud8"}, {\tt "le"}, {\tt
            "lu"}, {\tt "ld"}, {\tt "qe"}, {\tt "qu1"}, {\tt "qu8"},
          {\tt "qd1"}, {\tt "qd8"}, {\tt "ledq"}, {\tt "quqd1"}, {\tt
            "quqd8"}, {\tt "lequ1"}, {\tt "lequ3"}, {\tt "vv"}, {\tt
            "duq"}, {\tt "qqu"}, {\tt "qqq"}, {\tt "duu"} \}

}

\medskip

Similarly, \sfr takes as input bosonic dimension-8 operators from
Tables~\ref{tab:dim8-operators-higgs}, \ref{tab:dim8-operators-gauge},
\ref{tab:dim8-operators-gauge-higgs}, again rewritten here for
completeness from ref.~\cite{Murphy:2020rsh}.  For example, one can
use the following names in the list of dimension-8 operators:
\begin{align*}
 Q_{\varphi^4 D^4}^{(1)}   &\rightarrow {\tt ``phi4D4n1"}\\
 Q_{\varphi^6 \square}     &\rightarrow {\tt ``phi6Box"}\\
 Q_{G^2B^2}^{(4)}          &\rightarrow {\tt ``G2B2n4"}\\
 Q_{W^2B\varphi^2}^{(2)}   &\rightarrow {\tt ``W2Bphi2n2"}\\
 Q_{W^2\varphi^2D^2}^{(1)} &\rightarrow {\tt ``W2phi2D2n1"}\\
 ....  \nonumber
\end{align*}
%%%%%%%%%%%%%%%%%%%%%%%%%%%%%%%%%%%
Table~\ref{tab:dim8-operators-higgs} collects the pure Higgs
operators, i.e.\ operators constructed only out of the Higgs doublet,
$\varphi$, and covariant derivatives.  There, we performed a change of
basis in the operators of the $\varphi^{6} D^{2}$ class so that they
have immediate connection with the Warsaw basis.  The original
operators where defined in \cite{Murphy:2020rsh} as
\begin{align*}
    Q_{\varphi^6}^{(1)} & = (\varphi^{\dagger} \varphi)^2 (D_{\mu}
    \varphi^{\dagger} D^{\mu} \varphi) \,, \\
    Q_{\varphi^6}^{(2)} & = (\varphi^{\dagger} \varphi)
    (\varphi^{\dagger} \tau^I \varphi) (D_{\mu} \varphi^{\dagger}
    \tau^I D^{\mu} \varphi) \,,
%\numberthis
\end{align*}
and here we use instead the set
\begin{align*}
    Q_{\varphi^6 \square} &= (\varphi^\dagger \varphi)^{2} \square
    (\varphi^\dagger \varphi) \,, \\
    Q_{\varphi^{6} D^{2}} &= (\varphi^{\dagger} \varphi)
    (\varphi^\dagger D_\mu\varphi)^{\ast} (\varphi^\dagger
    D^\mu\varphi) \,,
%\numberthis
\end{align*}
which naturally extends the definition of the dimension $6$ operators
$Q_{\varphi\square}$ and $Q_{\varphi D}$ from
table~\ref{tab:sfr3-dim6}.  This change of basis is consistent with
the rest of the basis from ref.~\cite{Murphy:2020rsh}.  A proof of
this result can be found in appendix~F of ref.~\cite{LamprosPhD} for
any order in the EFT expansion.  Additionally, we added the number of
covariant derivatives in the naming of the operators that belong in
the third class, $\varphi^{4} D^{4}$, to avoid confusion with the SM
quartic Higgs operator, $\varphi^{4}$.

Table~\ref{tab:dim8-operators-gauge} collects the operators that are
constructed purely from gauge field strengths.  Therefore, each
operator there contains exactly four field strengths, and the operator
classes are further divided as $X^{4}$, where only one of the field
strengths of the $B$, $W$ or $G$ gauge fields appears in the operator,
$X^{3} X^{\prime}$, where the $G$ field strength appears thrice
together with a $B$ field strength in the operator, and finally
$X^{2}X^{\prime 2}$, where the operators are consisted of two pairs of
different field strengths.  The notation in this table follows exactly
ref.~\cite{Murphy:2020rsh}.  Finally,
table~\ref{tab:dim8-operators-gauge-higgs} collects the operators that
are constructed from a combination of Higgs doublets, $\varphi$, and
gauge field strengths.

The full list of names of bosonic dimension-8 operators in the basis
of ref.~\cite{Murphy:2020rsh} (with the modifications described above)
which can be included in \sfr v3 calculations reads as:

\medskip

{\small
  
\noindent {\tt OpList8} = \{ {\tt "phi8"}, {\tt "phi6Box"}, {\tt
  "phi6D2"}, {\tt "G2phi4n1"}, {\tt "G2phi4n2"}, {\tt "W2phi4n1"},
          {\tt "W2phi4n2"}, {\tt "W2phi4n3"}, {\tt "W2phi4n4"}, {\tt
            "WBphi4n1"}, {\tt "WBphi4n2"}, {\tt "B2phi4n1"}, {\tt
            "B2phi4n2"}, {\tt "G4n1"}, {\tt "G4n2"}, {\tt "G4n3"},
          {\tt "G4n4"}, {\tt "G4n5"}, {\tt "G4n6"}, {\tt "G4n7"}, {\tt
            "G4n8"}, {\tt "G4n9"}, {\tt "W4n1"}, {\tt "W4n2"}, {\tt
            "W4n3"}, {\tt "W4n4"}, {\tt "W4n5"}, {\tt "W4n6"}, {\tt
            "B4n1"}, {\tt "B4n2"}, {\tt "B4n3"}, {\tt "G3Bn1"}, {\tt
            "G3Bn2"}, {\tt "G3Bn3"}, {\tt "G3Bn4"}, {\tt "G2W2n1"},
          {\tt "G2W2n2"}, {\tt "G2W2n3"}, {\tt "G2W2n4"}, {\tt
            "G2W2n5"}, {\tt "G2W2n6"}, {\tt "G2W2n7"}, {\tt "G2B2n1"},
          {\tt "G2B2n2"}, {\tt "G2B2n3"}, {\tt "G2B2n4"}, {\tt
            "G2B2n5"}, {\tt "G2B2n6"}, {\tt "G2B2n7"}, {\tt "W2B2n1"},
          {\tt "W2B2n2"}, {\tt "W2B2n3"}, {\tt "W2B2n4"}, {\tt
            "W2B2n5"}, {\tt "W2B2n6"}, {\tt "W2B2n7"}, {\tt "phi4D4n1"},
          {\tt "phi4D4n2"}, {\tt "phi4D4n3"}, {\tt "G3phi2n1"}, {\tt
            "G3phi2n2"}, {\tt "W3phi2n1"}, {\tt "W3phi2n2"}, {\tt
            "W2Bphi2n1"}, \\ {\tt "W2Bphi2n2"}, {\tt "G2phi2D2n1"},
          {\tt "G2phi2D2n2"}, {\tt "G2phi2D2n3"}, {\tt "W2phi2D2n1"},
          {\tt "W2phi2D2n2"}, \\ {\tt "W2phi2D2n3"}, {\tt
            "W2phi2D2n4"}, {\tt "W2phi2D2n5"}, {\tt "W2phi2D2n6"},
          {\tt "WBphi2D2n1"}, {\tt "WBphi2D2n2"}, \\ {\tt
            "WBphi2D2n3"}, {\tt "WBphi2D2n4"}, {\tt "WBphi2D2n5"},
          {\tt "WBphi2D2n6"}, {\tt "B2phi2D2n1"}, {\tt "B2phi2D2n2"},
          \\ {\tt "B2phi2D2n3"}, {\tt "Wphi4D2n1"}, {\tt "Wphi4D2n2"},
             {\tt "Wphi4D2n3"}, {\tt "Wphi4D2n4"}, {\tt "Bphi4D2n1"},
             \\ {\tt "Bphi4D2n2"} \}

}

\newpage
%%%%%%%%%%%%%%%%%%%%%%%%%%%%%%%%%%%%%%%%%
\begin{table}[h!]
\centering
\small
\begin{adjustbox}{width=\textwidth,max width=\textwidth,max totalheight=\textheight,keepaspectratio}
\begin{tabular}{|c|c|c|c|c|c|} 
\hline
\multicolumn{2}{|c|}{$X^3$} & 
\multicolumn{2}{c|}{$\varphi^6$~ and~ $\varphi^4 D^2$} &
\multicolumn{2}{c|}{$\psi^2\varphi^3$}\\
\hline
$Q_G$                & $f^{ABC} G_\mu^{A\nu} G_\nu^{B\rho} G_\rho^{C\mu} $ &  
$Q_\varphi$       & $(\varphi^\dagger\varphi)^3$ &
$Q_{e\varphi}$           & $(\varphi^\dagger \varphi)(\bar l_p e_r \varphi)$\\
$Q_{\widetilde G}$          & $f^{ABC} \widetilde G_\mu^{A\nu} G_\nu^{B\rho} G_\rho^{C\mu} $ &   
$Q_{\varphi\Box}$ & $(\varphi^\dagger \varphi)\raisebox{-.5mm}{$\Box$}(\varphi^\dagger \varphi)$ &
$Q_{u\varphi}$           & $(\varphi^\dagger \varphi)(\bar q_p u_r \widetilde{\varphi})$\\
$Q_W$                & $\epsilon^{IJK} W_\mu^{I\nu} W_\nu^{J\rho} W_\rho^{K\mu}$ &    
$Q_{\varphi D}$   & $\left(\varphi^\dagger D^\mu\varphi\right)^{\ast} \left(\varphi^\dagger D_\mu\varphi\right)$ &
$Q_{d\varphi}$           & $(\varphi^\dagger \varphi)(\bar q_p d_r \varphi)$\\
$Q_{\widetilde W}$          & $\epsilon^{IJK} \widetilde W_\mu^{I\nu} W_\nu^{J\rho} W_\rho^{K\mu}$ &&&&\\    
\hline \hline
\multicolumn{2}{|c|}{$X^2\varphi^2$} &
\multicolumn{2}{c|}{$\psi^2 X\varphi$} &
\multicolumn{2}{c|}{$\psi^2\varphi^2 D$}\\ 
\hline
$Q_{\varphi G}$     & $\varphi^\dagger \varphi\, G^A_{\mu\nu} G^{A\mu\nu}$ & 
$Q_{eW}$               & $(\bar l_p \sigma^{\mu\nu} e_r) \tau^I \varphi W_{\mu\nu}^I$ &
$Q_{\varphi l}^{(1)}$      & $i(\varphi^{\dagger}\overset\leftrightarrow{D}_{\mu}\varphi)(\bar l_p \gamma^\mu l_r)$\\
$Q_{\varphi\widetilde G}$         & $\varphi^\dagger \varphi\, \widetilde G^A_{\mu\nu} G^{A\mu\nu}$ &  
$Q_{eB}$        & $(\bar l_p \sigma^{\mu\nu} e_r) \varphi B_{\mu\nu}$ &
$Q_{\varphi l}^{(3)}$      & $i(\varphi^{\dagger}\overset\leftrightarrow{D}_{\mu}\,\!\!\!^{I}\varphi)(\bar l_p \tau^I \gamma^\mu l_r)$\\
$Q_{\varphi W}$     & $\varphi^\dagger \varphi\, W^I_{\mu\nu} W^{I\mu\nu}$ & 
$Q_{uG}$        & $(\bar q_p \sigma^{\mu\nu} T^A u_r) \widetilde{\varphi}\, G_{\mu\nu}^A$ &
$Q_{\varphi e}$            & $i(\varphi^{\dagger}\overset\leftrightarrow{D}_{\mu}\varphi)(\bar e_p \gamma^\mu e_r)$\\
$Q_{\varphi\widetilde W}$         & $\varphi^\dagger \varphi\, \widetilde W^I_{\mu\nu} W^{I\mu\nu}$ &
$Q_{uW}$               & $(\bar q_p \sigma^{\mu\nu} u_r) \tau^I \widetilde{\varphi}\, W_{\mu\nu}^I$ &
$Q_{\varphi q}^{(1)}$      & $i(\varphi^{\dagger}\overset\leftrightarrow{D}_{\mu}\varphi)(\bar q_p \gamma^\mu q_r)$\\
$Q_{\varphi B}$     & $ \varphi^\dagger \varphi\, B_{\mu\nu} B^{\mu\nu}$ &
$Q_{uB}$        & $(\bar q_p \sigma^{\mu\nu} u_r) \widetilde{\varphi}\, B_{\mu\nu}$&
$Q_{\varphi q}^{(3)}$      & $i(\varphi^{\dagger}\overset\leftrightarrow{D}_{\mu}\,\!\!\!^{I}\varphi)(\bar q_p \tau^I \gamma^\mu q_r)$\\
$Q_{\varphi\widetilde B}$         & $\varphi^\dagger \varphi\, \widetilde B_{\mu\nu} B^{\mu\nu}$ &
$Q_{dG}$        & $(\bar q_p \sigma^{\mu\nu} T^A d_r) \varphi\, G_{\mu\nu}^A$ & 
$Q_{\varphi u}$            & $i(\varphi^{\dagger}\overset\leftrightarrow{D}_{\mu}\varphi)(\bar u_p \gamma^\mu u_r)$\\
$Q_{\varphi WB}$     & $ \varphi^\dagger \tau^I \varphi\, W^I_{\mu\nu} B^{\mu\nu}$ &
$Q_{dW}$               & $(\bar q_p \sigma^{\mu\nu} d_r) \tau^I \varphi\, W_{\mu\nu}^I$ &
$Q_{\varphi d}$            & $i(\varphi^{\dagger}\overset\leftrightarrow{D}_{\mu}\varphi)(\bar d_p \gamma^\mu d_r)$\\
$Q_{\varphi\widetilde WB}$ & $\varphi^\dagger \tau^I \varphi\, \widetilde W^I_{\mu\nu} B^{\mu\nu}$ &
$Q_{dB}$        & $(\bar q_p \sigma^{\mu\nu} d_r) \varphi\, B_{\mu\nu}$ &
$Q_{\varphi u d}$   & $i(\widetilde{\varphi}^\dagger D_\mu \varphi)(\bar u_p \gamma^\mu d_r)$\\
\hline
\end{tabular}
\end{adjustbox}
\\[0.4em]
\begin{adjustbox}{width=\textwidth,max width=\textwidth,max totalheight=\textheight,keepaspectratio}
\begin{tabular}{|c|c|c|c|c|c|}
\hline
\multicolumn{2}{|c|}{$(\bar LL)(\bar LL)$} & 
\multicolumn{2}{c|}{$(\bar RR)(\bar RR)$} &
\multicolumn{2}{c|}{$(\bar LL)(\bar RR)$}\\
\hline
$Q_{ll}$        & $(\bar l_p \gamma_\mu l_r)(\bar l_s \gamma^\mu l_t)$ &
$Q_{ee}$               & $(\bar e_p \gamma_\mu e_r)(\bar e_s \gamma^\mu e_t)$ &
$Q_{le}$               & $(\bar l_p \gamma_\mu l_r)(\bar e_s \gamma^\mu e_t)$ \\
$Q_{qq}^{(1)}$  & $(\bar q_p \gamma_\mu q_r)(\bar q_s \gamma^\mu q_t)$ &
$Q_{uu}$        & $(\bar u_p \gamma_\mu u_r)(\bar u_s \gamma^\mu u_t)$ &
$Q_{lu}$               & $(\bar l_p \gamma_\mu l_r)(\bar u_s \gamma^\mu u_t)$ \\
$Q_{qq}^{(3)}$  & $(\bar q_p \gamma_\mu \tau^I q_r)(\bar q_s \gamma^\mu \tau^I q_t)$ &
$Q_{dd}$        & $(\bar d_p \gamma_\mu d_r)(\bar d_s \gamma^\mu d_t)$ &
$Q_{ld}$               & $(\bar l_p \gamma_\mu l_r)(\bar d_s \gamma^\mu d_t)$ \\
$Q_{lq}^{(1)}$                & $(\bar l_p \gamma_\mu l_r)(\bar q_s \gamma^\mu q_t)$ &
$Q_{eu}$                      & $(\bar e_p \gamma_\mu e_r)(\bar u_s \gamma^\mu u_t)$ &
$Q_{qe}$               & $(\bar q_p \gamma_\mu q_r)(\bar e_s \gamma^\mu e_t)$ \\
$Q_{lq}^{(3)}$                & $(\bar l_p \gamma_\mu \tau^I l_r)(\bar q_s \gamma^\mu \tau^I q_t)$ &
$Q_{ed}$                      & $(\bar e_p \gamma_\mu e_r)(\bar d_s\gamma^\mu d_t)$ &
$Q_{qu}^{(1)}$         & $(\bar q_p \gamma_\mu q_r)(\bar u_s \gamma^\mu u_t)$ \\ 
&& 
$Q_{ud}^{(1)}$                & $(\bar u_p \gamma_\mu u_r)(\bar d_s \gamma^\mu d_t)$ &
$Q_{qu}^{(8)}$         & $(\bar q_p \gamma_\mu T^A q_r)(\bar u_s \gamma^\mu T^A u_t)$ \\ 
&& 
$Q_{ud}^{(8)}$                & $(\bar u_p \gamma_\mu T^A u_r)(\bar d_s \gamma^\mu T^A d_t)$ &
$Q_{qd}^{(1)}$ & $(\bar q_p \gamma_\mu q_r)(\bar d_s \gamma^\mu d_t)$ \\
&&&&
$Q_{qd}^{(8)}$ & $(\bar q_p \gamma_\mu T^A q_r)(\bar d_s \gamma^\mu T^A d_t)$\\
\hline \hline
\multicolumn{2}{|c|}{$(\bar LR)(\bar RL)$ and $(\bar LR)(\bar LR)$} &
\multicolumn{4}{c|}{$B$-violating}\\\hline
$Q_{ledq}$ & $(\bar l_p^j e_r)(\bar d_s q_t^j)$ &
$Q_{duq}$ & \multicolumn{3}{c|}{$\epsilon ^{\alpha\beta\gamma} \epsilon _{jk} 
 \left[ (d^\alpha_p)^T C u^\beta_r \right]\left[(q^{\gamma j}_s)^T C l^k_t\right]$}\\
$Q_{quqd}^{(1)}$ & $(\bar q_p^j u_r) \epsilon _{jk} (\bar q_s^k d_t)$ &
$Q_{qqu}$ & \multicolumn{3}{c|}{$\epsilon ^{\alpha\beta\gamma} \epsilon _{jk} 
  \left[ (q^{\alpha j}_p)^T C q^{\beta k}_r \right]\left[(u^\gamma_s)^T C e_t\right]$}\\
$Q_{quqd}^{(8)}$ & $(\bar q_p^j T^A u_r) \epsilon _{jk} (\bar q_s^k T^A d_t)$ &
$Q_{qqq}$ & \multicolumn{3}{c|}{$\epsilon ^{\alpha\beta\gamma} \epsilon _{jn} \epsilon _{km} 
  \left[ (q^{\alpha j}_p)^T C q^{\beta k}_r \right]\left[(q^{\gamma m}_s)^T C l^n_t\right]$}\\
$Q_{lequ}^{(1)}$ & $(\bar l_p^j e_r) \epsilon _{jk} (\bar q_s^k u_t)$ &
$Q_{duu}$ & \multicolumn{3}{c|}{$\epsilon ^{\alpha\beta\gamma} 
  \left[ (d^\alpha_p)^T C u^\beta_r \right]\left[(u^\gamma_s)^T C e_t\right]$}\\
$Q_{lequ}^{(3)}$ & $(\bar l_p^j \sigma_{\mu\nu} e_r) \epsilon _{jk} (\bar q_s^k \sigma^{\mu\nu} u_t)$ &
& \multicolumn{3}{c|}{}\\
\hline
\end{tabular}
\end{adjustbox}
\caption[Dimension $6$ operators]{The full set of dimension $6$
  operators in Warsaw basis \cite{Grzadkowski:2010es}.  The sub-tables
  in the two upper rows collect all operators except the four-fermion
  ones, which are collected separately in the sub-tables of the two
  bottom rows.}
\label{tab:sfr3-dim6}
\end{table}

\newpage

%\section{Dimension 8 operators}
%
\begin{table}[h!]
\centering
\small
\begin{adjustbox}{max width=\textwidth,max totalheight=\textheight,keepaspectratio,center}
\begin{tabular}{|c|c|c|c|c|c|}
\hline
\multicolumn{2}{|c|}{$\varphi^8$} &
\multicolumn{2}{c|}{$\varphi^6D^2$} &
\multicolumn{2}{c|}{$\varphi^4D^4$} \\
\hline
$Q_{\varphi^8}$  &  $(\varphi^\dagger \varphi)^4$ &
$Q_{\varphi^6 \square}$ &
$(\varphi^\dagger \varphi)^{2} \square (\varphi^\dagger \varphi)$ &
$Q_{\varphi^4 D^4}^{(1)}$  &
$(D_{\mu} \varphi^{\dagger} D_{\nu} \varphi) (D^{\nu} \varphi^{\dagger} D^{\mu} \varphi)$ \\
& & $Q_{\varphi^{6} D^{2}}$ & $(\varphi^{\dagger} \varphi)
(\varphi^\dagger D_\mu\varphi)^{\ast} (\varphi^\dagger D^\mu\varphi)$  &
$Q_{\varphi^4 D^4}^{(2)}$  &
$(D_{\mu} \varphi^{\dagger} D_{\nu} \varphi) (D^{\mu} \varphi^{\dagger} D^{\nu} \varphi)$ \\
& & & & $Q_{\varphi^4 D^4}^{(3)}$  &
$(D_{\mu} \varphi^{\dagger} D^{\mu} \varphi) (D_{\nu} \varphi^{\dagger} D^{\nu} \varphi)$ \\
\hline
\end{tabular}
\end{adjustbox}
\caption{Dimension $8$ operators containing only the Higgs field.
  Table taken from ref.~\cite{Murphy:2020rsh} except for the two
  operators in $\varphi^6 D^2$ class that have been modified as
  discussed in this Appendix.}
\label{tab:dim8-operators-higgs}
\end{table}
\begin{table}[h!]
\centering
\small
\begin{adjustbox}{max width=\textwidth,max totalheight=\textheight,keepaspectratio,center}
\begin{tabular}{|c|c|c|c|}
\hline
\multicolumn{2}{|c|}{$X^4,\; X^3 X^{\prime}$} &
\multicolumn{2}{c|}{$X^2 X^{\prime 2}$} \\
\hline
$Q_{G^4}^{(1)}$  &  $(G_{\mu\nu}^A G^{A\mu\nu}) (G_{\rho\sigma}^B G^{B\rho\sigma})$ &
$Q_{G^2W^2}^{(1)}$  &  $(W_{\mu\nu}^I W^{I\mu\nu}) (G_{\rho\sigma}^A G^{A\rho\sigma})$ \\
$Q_{G^4}^{(2)}$  &  $(G_{\mu\nu}^A \widetilde{G}^{A\mu\nu}) (G_{\rho\sigma}^B \widetilde{G}^{B\rho\sigma})$ &
$Q_{G^2W^2}^{(2)}$  &  $(W_{\mu\nu}^I \widetilde{W}^{I\mu\nu}) (G_{\rho\sigma}^A \widetilde{G}^{A\rho\sigma})$ \\
$Q_{G^4}^{(3)}$  &  $(G_{\mu\nu}^A G^{B\mu\nu}) (G_{\rho\sigma}^A G^{B\rho\sigma})$ &
$Q_{G^2W^2}^{(3)}$  &  $(W_{\mu\nu}^I G^{A\mu\nu}) (W_{\rho\sigma}^I G^{A\rho\sigma})$ \\
$Q_{G^4}^{(4)}$  &  $(G_{\mu\nu}^A \widetilde{G}^{B\mu\nu}) (G_{\rho\sigma}^A \widetilde{G}^{B\rho\sigma})$ &
$Q_{G^2W^2}^{(4)}$  &  $(W_{\mu\nu}^I \widetilde{G}^{A\mu\nu}) (W_{\rho\sigma}^I \widetilde{G}^{A\rho\sigma})$ \\
$Q_{G^4}^{(5)}$  &  $(G_{\mu\nu}^A G^{A\mu\nu}) (G_{\rho\sigma}^B \widetilde{G}^{B\rho\sigma})$ &
$Q_{G^2W^2}^{(5)}$  &  $(W_{\mu\nu}^I \widetilde{W}^{I\mu\nu}) (G_{\rho\sigma}^A G^{A\rho\sigma})$ \\
$Q_{G^4}^{(6)}$  &  $(G_{\mu\nu}^A G^{B\mu\nu}) (G_{\rho\sigma}^A \widetilde{G}^{B\rho\sigma})$ &
$Q_{G^2W^2}^{(6)}$  &  $(W_{\mu\nu}^I W^{I\mu\nu}) (G_{\rho\sigma}^A \widetilde{G}^{A\rho\sigma})$ \\
$Q_{G^4}^{(7)}$  &  $d^{ABE} d^{CDE} (G_{\mu\nu}^A G^{B\mu\nu}) (G_{\rho\sigma}^C G^{D\rho\sigma})$ &
$Q_{G^2W^2}^{(7)}$  &  $(W_{\mu\nu}^I G^{A\mu\nu}) (W_{\rho\sigma}^I \widetilde{G}^{A\rho\sigma})$ \\
$Q_{G^4}^{(8)}$  &  $d^{ABE} d^{CDE} (G_{\mu\nu}^A \widetilde{G}^{B\mu\nu}) (G_{\rho\sigma}^C \widetilde{G}^{D\rho\sigma})$ &
$Q_{G^2B^2}^{(1)}$  &  $(B_{\mu\nu} B^{\mu\nu}) (G_{\rho\sigma}^A G^{A\rho\sigma})$ \\
$Q_{G^4}^{(9)}$  &  $d^{ABE} d^{CDE} (G_{\mu\nu}^A G^{B\mu\nu}) (G_{\rho\sigma}^C \widetilde{G}^{D\rho\sigma})$ &
$Q_{G^2B^2}^{(2)}$  &  $(B_{\mu\nu} \widetilde{B}^{\mu\nu}) (G_{\rho\sigma}^A \widetilde{G}^{A\rho\sigma})$ \\
$Q_{W^4}^{(1)}$  &  $(W_{\mu\nu}^I W^{I\mu\nu}) (W_{\rho\sigma}^J W^{J\rho\sigma})$ &
$Q_{G^2B^2}^{(3)}$  &  $(B_{\mu\nu} G^{A\mu\nu}) (B_{\rho\sigma} G^{A\rho\sigma})$ \\
$Q_{W^4}^{(2)}$  &  $(W_{\mu\nu}^I \widetilde{W}^{I\mu\nu}) (W_{\rho\sigma}^J \widetilde{W}^{J\rho\sigma})$ &
$Q_{G^2B^2}^{(4)}$  &  $(B_{\mu\nu} \widetilde{G}^{A\mu\nu}) (B_{\rho\sigma} \widetilde{G}^{A\rho\sigma})$ \\
$Q_{W^4}^{(3)}$  &  $(W_{\mu\nu}^I W^{J\mu\nu}) (W_{\rho\sigma}^I W^{J\rho\sigma})$ &
$Q_{G^2B^2}^{(5)}$  &  $(B_{\mu\nu} \widetilde{B}^{\mu\nu}) (G_{\rho\sigma}^A G^{A\rho\sigma})$ \\
$Q_{W^4}^{(4)}$  &  $(W_{\mu\nu}^I \widetilde{W}^{J\mu\nu}) (W_{\rho\sigma}^I \widetilde{W}^{J\rho\sigma})$ &
$Q_{G^2B^2}^{(6)}$  &  $(B_{\mu\nu} B^{\mu\nu}) (G_{\rho\sigma}^A \widetilde{G}^{A\rho\sigma})$ \\
$Q_{W^4}^{(5)}$  &  $(W_{\mu\nu}^I W^{I\mu\nu}) (W_{\rho\sigma}^J \widetilde{W}^{J\rho\sigma})$ &
$Q_{G^2B^2}^{(7)}$  &  $(B_{\mu\nu} G^{A\mu\nu}) (B_{\rho\sigma} \widetilde{G}^{A\rho\sigma})$ \\
$Q_{W^4}^{(6)}$  &  $(W_{\mu\nu}^I W^{J\mu\nu}) (W_{\rho\sigma}^I \widetilde{W}^{J\rho\sigma})$ &
$Q_{W^2B^2}^{(1)}$  &  $(B_{\mu\nu} B^{\mu\nu}) (W_{\rho\sigma}^I W^{I\rho\sigma})$ \\
$Q_{B^4}^{(1)}$  &  $(B_{\mu\nu} B^{\mu\nu}) (B_{\rho\sigma} B^{\rho\sigma})$ &
$Q_{W^2B^2}^{(2)}$  &  $(B_{\mu\nu} \widetilde{B}^{\mu\nu}) (W_{\rho\sigma}^I \widetilde{W}^{I\rho\sigma})$ \\
$Q_{B^4}^{(2)}$  &  $(B_{\mu\nu} \widetilde{B}^{\mu\nu}) (B_{\rho\sigma} \widetilde{B}^{\rho\sigma})$ &
$Q_{W^2B^2}^{(3)}$  &  $(B_{\mu\nu} W^{I\mu\nu}) (B_{\rho\sigma} W^{I\rho\sigma})$ \\
$Q_{B^4}^{(3)}$  &  $(B_{\mu\nu} B^{\mu\nu}) (B_{\rho\sigma} \widetilde{B}^{\rho\sigma})$ &
$Q_{W^2B^2}^{(4)}$  &  $(B_{\mu\nu} \widetilde{W}^{I\mu\nu}) (B_{\rho\sigma} \widetilde{W}^{I\rho\sigma})$ \\
$Q_{G^3B}^{(1)}$  &  $d^{ABC} (B_{\mu\nu} G^{A\mu\nu}) (G_{\rho\sigma}^B G^{C\rho\sigma})$ &
$Q_{W^2B^2}^{(5)}$  &  $(B_{\mu\nu} \widetilde{B}^{\mu\nu}) (W_{\rho\sigma}^I W^{I\rho\sigma})$ \\
$Q_{G^3B}^{(2)}$  &  $d^{ABC} (B_{\mu\nu} \widetilde{G}^{A\mu\nu}) (G_{\rho\sigma}^B \widetilde{G}^{C\rho\sigma})$ &
$Q_{W^2B^2}^{(6)}$  &  $(B_{\mu\nu} B^{\mu\nu}) (W_{\rho\sigma}^I \widetilde{W}^{I\rho\sigma})$ \\
$Q_{G^3B}^{(3)}$  &  $d^{ABC} (B_{\mu\nu} \widetilde{G}^{A\mu\nu}) (G_{\rho\sigma}^B G^{C\rho\sigma})$ &
$Q_{W^2B^2}^{(7)}$  &  $(B_{\mu\nu} W^{I\mu\nu}) (B_{\rho\sigma} \widetilde{W}^{I\rho\sigma})$ \\
$Q_{G^3B}^{(4)}$  &  $d^{ABC} (B_{\mu\nu} G^{A\mu\nu}) (G_{\rho\sigma}^B \widetilde{G}^{C\rho\sigma})$ & & \\
\hline
\end{tabular}
\end{adjustbox}
\caption{Dimension $8$ operators containing only gauge field
  strengths.  Table taken from ref.~\cite{Murphy:2020rsh}.}
\label{tab:dim8-operators-gauge}
\end{table}
\begin{table}[h!]
\centering
\small
\begin{adjustbox}{max width=\textwidth,max totalheight=\textheight,keepaspectratio,center}
\begin{tabular}{|c|c|c|c|}
\hline
\multicolumn{2}{|c|}{$X^3\varphi^2$} &
\multicolumn{2}{c|}{$X^2\varphi^4$} \\
\hline
$Q_{G^3\varphi^2}^{(1)}$  &  $f^{ABC} (\varphi^\dag \varphi) G_{\mu}^{A\nu} G_{\nu}^{B\rho} G_{\rho}^{C\mu}$ &
$Q_{G^2\varphi^4}^{(1)}$  & $(\varphi^\dag \varphi)^2 G^A_{\mu\nu} G^{A\mu\nu}$ \\
$Q_{G^3\varphi^2}^{(2)}$  &  $f^{ABC} (\varphi^\dag \varphi) G_{\mu}^{A\nu} G_{\nu}^{B\rho} \widetilde{G}_{\rho}^{C\mu}$ &
$Q_{G^2\varphi^4}^{(2)}$  & $(\varphi^\dag \varphi)^2 \widetilde G^A_{\mu\nu} G^{A\mu\nu}$ \\
$Q_{W^3\varphi^2}^{(1)}$  &  $\epsilon^{IJK} (\varphi^\dag \varphi) W_{\mu}^{I\nu} W_{\nu}^{J\rho} W_{\rho}^{K\mu}$ &
$Q_{W^2\varphi^4}^{(1)}$  & $(\varphi^\dag \varphi)^2 W^I_{\mu\nu} W^{I\mu\nu}$ \\
$Q_{W^3\varphi^2}^{(2)}$  &  $\epsilon^{IJK} (\varphi^\dag \varphi) W_{\mu}^{I\nu} W_{\nu}^{J\rho} \widetilde{W}_{\rho}^{K\mu}$ &
$Q_{W^2\varphi^4}^{(2)}$  & $(\varphi^\dag \varphi)^2 \widetilde W^I_{\mu\nu} W^{I\mu\nu}$ \\
$Q_{W^2B\varphi^2}^{(1)}$  &  $\epsilon^{IJK} (\varphi^\dag \tau^I \varphi) B_{\mu}^{\,\nu} W_{\nu}^{J\rho} W_{\rho}^{K\mu}$ &
$Q_{W^2\varphi^4}^{(3)}$  & $(\varphi^\dag \tau^I \varphi) (\varphi^\dag \tau^J \varphi) W^I_{\mu\nu} W^{J\mu\nu}$ \\
$Q_{W^2B\varphi^2}^{(2)}$  &  $\epsilon^{IJK} (\varphi^\dag \tau^I \varphi) (\widetilde{B}^{\mu\nu} W_{\nu\rho}^J W_{\mu}^{K\rho} + B^{\mu\nu} W_{\nu\rho}^J \widetilde{W}_{\mu}^{K\rho})$ &
$Q_{W^2\varphi^4}^{(4)}$  & $(\varphi^\dag \tau^I \varphi) (\varphi^\dag \tau^J \varphi) \widetilde W^I_{\mu\nu} W^{J\mu\nu}$ \\
& &
$Q_{WB\varphi^4}^{(1)}$  & $ (\varphi^\dag \varphi) (\varphi^\dag \tau^I \varphi) W^I_{\mu\nu} B^{\mu\nu}$ \\
& &
$Q_{WB\varphi^4}^{(2)}$  & $(\varphi^\dag \varphi) (\varphi^\dag \tau^I \varphi) \widetilde W^I_{\mu\nu} B^{\mu\nu}$ \\
& &
$Q_{B^2\varphi^4}^{(1)}$  & $ (\varphi^\dag \varphi)^2 B_{\mu\nu} B^{\mu\nu}$ \\
& &
$Q_{B^2\varphi^4}^{(2)}$  & $(\varphi^\dag \varphi)^2 \widetilde B_{\mu\nu} B^{\mu\nu}$ \\
%%%%%%%%%%%%
\hline
\hline
%%%%%%%%%%%%
\multicolumn{2}{|c|}{$X^2\varphi^2D^2$} &
\multicolumn{2}{c|}{$X\varphi^4D^2$} \\
\hline
$Q_{G^2\varphi^2D^2}^{(1)}$  &  $(D^{\mu} \varphi^{\dag} D^{\nu} \varphi) G_{\mu\rho}^A G_{\nu}^{A \rho}$ &
$Q_{W\varphi^4D^2}^{(1)}$  & $(\varphi^{\dag} \varphi) (D^{\mu} \varphi^{\dag} \tau^I D^{\nu} \varphi) W_{\mu\nu}^I$ \\
$Q_{G^2\varphi^2D^2}^{(2)}$  &  $(D^{\mu} \varphi^{\dag} D_{\mu} \varphi) G_{\nu\rho}^A G^{A \nu\rho}$ &
$Q_{W\varphi^4D^2}^{(2)}$  & $(\varphi^{\dag} \varphi) (D^{\mu} \varphi^{\dag} \tau^I D^{\nu} \varphi) \widetilde{W}_{\mu\nu}^I$ \\
$Q_{G^2\varphi^2D^2}^{(3)}$  &  $(D^{\mu} \varphi^{\dag} D_{\mu} \varphi) G_{\nu\rho}^A \widetilde{G}^{A \nu\rho}$ &
$Q_{W\varphi^4D^2}^{(3)}$  & $\epsilon^{IJK} (\varphi^{\dag} \tau^I \varphi) (D^{\mu} \varphi^{\dag} \tau^J D^{\nu} \varphi) W_{\mu\nu}^K$ \\
$Q_{W^2\varphi^2D^2}^{(1)}$  &  $(D^{\mu} \varphi^{\dag} D^{\nu} \varphi) W_{\mu\rho}^I W_{\nu}^{I \rho}$ &
$Q_{W\varphi^4D^2}^{(4)}$  & $\epsilon^{IJK} (\varphi^{\dag} \tau^I \varphi) (D^{\mu} \varphi^{\dag} \tau^J D^{\nu} \varphi) \widetilde{W}_{\mu\nu}^K$ \\
$Q_{W^2\varphi^2D^2}^{(2)}$  &  $(D^{\mu} \varphi^{\dag} D_{\mu} \varphi) W_{\nu\rho}^I W^{I \nu\rho}$ &
$Q_{B\varphi^4D^2}^{(1)}$  & $(\varphi^{\dag} \varphi) (D^{\mu} \varphi^{\dag} D^{\nu} \varphi) B_{\mu\nu}$ \\
$Q_{W^2\varphi^2D^2}^{(3)}$  &  $(D^{\mu} \varphi^{\dag} D_{\mu} \varphi) W_{\nu\rho}^I \widetilde{W}^{I \nu\rho}$ &
$Q_{B\varphi^4D^2}^{(2)}$  & $(\varphi^{\dag} \varphi) (D^{\mu} \varphi^{\dag} D^{\nu} \varphi) \widetilde{B}_{\mu\nu}$ \\
$Q_{W^2\varphi^2D^2}^{(4)}$  &  $i \epsilon^{IJK} (D^{\mu} \varphi^{\dag} \tau^I D^{\nu} \varphi) W_{\mu\rho}^J W_{\nu}^{K \rho}$ &
                       & \\
$Q_{W^2\varphi^2D^2}^{(5)}$  &  $\epsilon^{IJK} (D^{\mu} \varphi^{\dag} \tau^I D^{\nu} \varphi) (W_{\mu\rho}^J \widetilde{W}_{\nu}^{K \rho} - \widetilde{W}_{\mu\rho}^J W_{\nu}^{K \rho})$ &
                       & \\
$Q_{W^2\varphi^2D^2}^{(6)}$  &  $i \epsilon^{IJK} (D^{\mu} \varphi^{\dag} \tau^I D^{\nu} \varphi) (W_{\mu\rho}^J \widetilde{W}_{\nu}^{K \rho} + \widetilde{W}_{\mu\rho}^J W_{\nu}^{K \rho})$ &
                       & \\
$Q_{WB\varphi^2D^2}^{(1)}$  &  $(D^{\mu} \varphi^{\dag} \tau^I D_{\mu} \varphi) B_{\nu\rho} W^{I \nu\rho}$ &
                       & \\
$Q_{WB\varphi^2D^2}^{(2)}$  &  $(D^{\mu} \varphi^{\dag} \tau^I D_{\mu} \varphi) B_{\nu\rho} \widetilde{W}^{I \nu\rho}$ &
                       & \\
$Q_{WB\varphi^2D^2}^{(3)}$  &  $i (D^{\mu} \varphi^{\dag} \tau^I D^{\nu} \varphi) (B_{\mu\rho} W_{\nu}^{I \rho} - B_{\nu\rho} W_{\mu}^{I\rho})$ &
                       & \\
$Q_{WB\varphi^2D^2}^{(4)}$  &  $(D^{\mu} \varphi^{\dag} \tau^I D^{\nu} \varphi) (B_{\mu\rho} W_{\nu}^{I \rho} + B_{\nu\rho} W_{\mu}^{I\rho})$ &
                       & \\
$Q_{WB\varphi^2D^2}^{(5)}$  &  $i (D^{\mu} \varphi^{\dag} \tau^I D^{\nu} \varphi) (B_{\mu\rho} \widetilde{W}_\nu^{^I \rho} - B_{\nu\rho} \widetilde{W}_\mu^{^I \rho})$ &
                       & \\
$Q_{WB\varphi^2D^2}^{(6)}$  &  $(D^{\mu} \varphi^{\dag} \tau^I D^{\nu} \varphi) (B_{\mu\rho} \widetilde{W}_\nu^{^I \rho} + B_{\nu\rho} \widetilde{W}_\mu^{^I \rho})$ &
                       & \\
$Q_{B^2\varphi^2D^2}^{(1)}$  &  $(D^{\mu} \varphi^{\dag} D^{\nu} \varphi) B_{\mu\rho} B_{\nu}^{\,\,\,\rho}$ &
                       & \\
$Q_{B^2\varphi^2D^2}^{(2)}$  &  $(D^{\mu} \varphi^{\dag} D_{\mu} \varphi) B_{\nu\rho} B^{\nu\rho}$ &
                       & \\
$Q_{B^2\varphi^2D^2}^{(3)}$  &  $(D^{\mu} \varphi^{\dag} D_{\mu} \varphi) B_{\nu\rho} \widetilde{B}^{\nu\rho}$ &
                       & \\
%%%%%%%%%%%%
\hline
\end{tabular}
%%%%%%%
\end{adjustbox}
\caption{Dimension $8$ operators containing both gauge field strengths
  and the Higgs field.  Table taken (and modified according to our
  notation) from ref.~\cite{Murphy:2020rsh}.}
\label{tab:dim8-operators-gauge-higgs}
\end{table}

\newpage
\hskip 1cm
%%%%%%%%%%% BIBLIOGRAPHY %%%%%%%%%%%%%%%%%
\bibliographystyle{elsarticle-num}
\bibliography{EFT}{}

\end{document}